\documentclass[usenatbib, backref,breaklinks,colorlinks,citecolor=blue]{mnras}
\usepackage{natbib}
\usepackage{amssymb}
\usepackage{amsmath}
\usepackage{graphicx}
\usepackage[normalem]{ulem}

\providecommand{\grad}{\ensuremath{g_{\mathrm{rad},s}}}
\providecommand{\gradun}{\ensuremath{g_\mathrm{rad, CAK}}}
\providecommand{\geff}{\ensuremath{g_{\mathrm{eff},s}}}
\providecommand{\lamcc}{\ensuremath{\Lambda_\mathrm{cc}}}
\providecommand{\lamrc}{\ensuremath{\Lambda_\mathrm{rc}}}
\providecommand{\scrit}{\ensuremath{s_\mathrm{crit}}}
\providecommand{\rcrit}{\ensuremath{r_c}}
\providecommand{\rcritz}{\ensuremath{r_\mathrm{c, \theta=0}}}
\providecommand{\Mcak}{\ensuremath{\dot M_\mathrm{CAK}}}
\providecommand{\mcak}{\ensuremath{\dot m_\mathrm{CAK}}}
\providecommand{\mtz}{\ensuremath{\dot m_\mathrm{\theta=0}}}
\providecommand{\mdip}{\ensuremath{\dot m_\mathrm{dip}}}
\providecommand{\mdipsat}{\ensuremath{\dot m_\mathrm{dip,OTC}}}
\providecommand{\mtzsat}{\ensuremath{\dot m_\mathrm{gen,\theta=0}}}
\providecommand{\Mglo}{\ensuremath{\dot M_\mathrm{global}}}
\providecommand{\gammael}{\ensuremath{\Gamma_\mathrm{el}}}
\providecommand{\sigsat}{\ensuremath{\Sigma}}
\providecommand{\sigsatz}{\ensuremath{\Sigma_\mathrm{0}}}
\providecommand{\sigsatrot}{\ensuremath{\Sigma_\mathrm{rot}}}
\providecommand{\sigsatrotm}{\ensuremath{\Sigma_\mathrm{rot,m}}}

\newcommand{\unit}[1]{\ensuremath{\, \mathrm{#1}}}
\newcommand{\e}[1]{\ensuremath{\times 10^{#1}}}

\title[Magnetic Massive Star Winds I]{Effect of a magnetic field on massive star winds I: mass-loss and velocity for a dipole field}
\author[C. Bard, R.H.D. Townsend]{Christopher Bard,$^{1}$\thanks{email: bard@astro.wisc.edu} and Richard H. D. Townsend$^{1}$\\
$^{1}$Department of Astronomy, University of Wisconsin-Madison, 475 N. Charter St., Madison, WI, 53706, USA}

\date{Accepted XXX. Received YYY; in original form ZZZ}
\pubyear{2016}

\begin{document}
\label{firstpage}
\pagerange{\pageref{firstpage}--\pageref{lastpage}}
\maketitle

\begin{abstract}
We generalize the Rigid-Field Hydrodynamic equations to accommodate arbitrary magnetic field topologies, resulting in a new Arbitrary Rigid-Field Hydrodynamic (ARFHD) formalism.
We undertake a critical point calculation of the steady-state ARFHD equations with a CAK-type radiative acceleration and determine the effects of a dipole magnetic field on the usual CAK mass-loss rate and velocity structure. 
Enforcing the proper optically-thin limit for the radiative line-acceleration is found to decrease both the mass-loss and wind acceleration, while rotation boosts both properties.
We define optically-thin-correction and rotation parameters to quantify these effects on the global mass-loss rate and develop scaling laws for the surface mass-flux as a function of surface colatitude.
These scaling laws are found to agree with previous laws derived from magnetohydrodynamic simulations of magnetospheres.
The dipole magnetosphere velocity structure is found to differ from a global beta-velocity law, which contradicts a central assumption of the previously-developed XADM model of X-ray emission from magnetospheres.
\end{abstract}


\section{Introduction}
\label{sec:Intro}
In the last decade, spectropolarimetric surveys of OB stars have revealed that about 5-10\% of these massive stars have large-scale, organized magnetic fields (MiMeS: \citealp{wade14}; BOB: \citealp{morel15}).
Such detectable magnetic fields ($B\gtrsim 100\unit{G}$) have a significant effect on the stellar wind, both channelling and trapping plasma within a stellar magnetosphere.
This accumulated plasma produces extrastellar emission in optical (e.g. \citealp{howarth07}, \citealp{bohlender11}, \citealp{grunhut12} and references therein), infrared \citep{eikenberry14}, radio \citep{linsky92, chandra15}, and X-ray \citep{naze14, naze15}.
Furthermore, this emission exhibits a rotational modulation as the plasma is forced by the magnetic field to co-rotate with the star.

Similar advances in magnetosphere theory have also followed, starting with the pioneering magnetohydrodynamics (MHD) simulations of \citet{uddoula02}.
They developed a ``wind magnetic confinement parameter'' to characterize the interplay between the stellar magnetic field and flow:
\begin{align}
\label{eq:eta_star}\eta_*\equiv \frac{B_\mathrm{eq}^2 R_*^2}{\dot M_{B=0} v_{\infty,B=0}},
\end{align}
with $\dot M_{B=0}$ and $v_{\infty,B=0}$ being the stellar mass-loss rate and terminal velocity if the star had no magnetic field.

The confinement parameter $\eta_*$ has become the canonical value adopted in scaling relations to explain the size \citep{uddoula02}, the mass-loss \citep{uddoula08}, the spin-down \citep{uddoula09}, and, with the critical rotation fraction $\omega$, the classification \citep{petit13} of magnetospheres.
However, $\eta_*$ itself depends on non-magnetic values, ignoring any effects of the magnetic field.
How does the magnetic field change the mass-loss rate and velocity?
Can we use these new values to make a better confinement parameter?

Traditionally, $\dot M$ and $v_\infty$ have been determined by analyzing the equation of motion for a line-driven wind (\citealp{cak75}; hereafter CAK) and solving for the so-called ``critical point''.
Over the years, various modifications to the base CAK model (finite-disk effect: \citealp{friend86}, \citealp{pauldrach86}; depth-dependent force multiplier parameters: \citealp{kudritzki02}) have led to more realistic predictions of the mass-loss and terminal velocities.
Other methods have been developed to improve on these estimates, such as a Monte Carlo method \citep{vink00, noebauer15} and a scattering source function technique \citep{sundqvist15}.
For now, we use the CAK line-driving force in order to take the first steps towards understanding the effect of a dipole field on a stellar wind.

In this paper, we present and study the Arbitrary Rigid-Field Hydrodynamics (ARFHD) equations, an extension of Rigid-Field Hydrodynamics (RFHD) \citep{townsend07} to account for non-dipole magnetic geometries (though we will consider only dipolar topologies in this analysis).
RFHD was originally developed as an extension of the Rigidly Rotating Magnetosphere (RRM) model \citep{townsend05} for centrifugal magnetospheres, whose large magnetic fields make MHD simulations very impractical.
In this ansatz, the magnetic fields are assumed to be completely rigid ($\eta_*\to\infty$), channeling the stellar wind along quasi-one-dimensional flux tubes.
This allows each field line to be studied and simulated independently from one another, though this does miss important multi-dimensional effects present in the MHD simulations.
In essence, the MHD studies approach the subject of massive-star magnetospheres from the regime of low magnetic confinement; ARFHD approaches this subject from the opposite regime of strong magnetic confinement.
By blending both studies, we can set limits on the behavior of magnetospheres.

In Section \ref{sec:hydro}, we present the reformulated ARFHD equations and define all the terms, including external sources of acceleration and cooling. Following this, we develop the critical point equations for an arbitrary magnetic configuration in Section \ref{sec:topo} and an algorithm for determing the critical point location in Section \ref{sec:crit}. 
Section \ref{sec:dip_topo} details the implementation and application of an aligned magnetic dipole radiation-driven wind model which includes the effect of stellar rotation.
We present analytic scalings of the surface mass-flux in Section \ref{sec:mass_flux} and model results for the critical point location (Section \ref{sec:rcrit}), velocity structure (Section \ref{sec:term_veloc}), and, finally, the global mass-loss rate (Section \ref{sec:mass_loss}).

\section{Arbitrary Rigid-Field Hydrodynamic Equations}
\label{sec:hydro}
Following \citet{townsend07} (hereafter T07), we extend the Rigid-Field Hydrodynamics (RFHD) model to incorporate arbitrary magnetic field line configurations, creating an Arbitrary Rigid-Field Hydrodynamics (ARFHD) model.
In this section, we recap the key assumptions and equations of RFHD, with additional commentary pertaining to ARFHD when relevant.

In the model, the key assumption is that the magnetic field is sufficiently strong to be effectively rigid (corresponding to $\eta_* \rightarrow \infty$).
This rigid field is tethered to the star and co-rotates with it.
Additionally, since the magnetic Reynolds number in the magnetosphere is so large ($\sim 10^{15}$), we assume that the ``frozen flux'' condition of ideal MHD applies.
As a result, the stiff magnetic field channels magnetospheric plasma flows along the field lines.
These trajectories are pre-determined from the chosen stellar magnetic topology, though the plasma state (density, velocity, temperature, etc.) is determined by the hydrodynamics of the flow along each magnetic field line.

The field lines are approximated as quasi-one-dimensional flux tubes, with ``quasi-'' referring to their varying cross-sectional area.
Under the requirement that local magnetic flux is conserved ($\nabla\cdot \bf{B}=0$), the cross-sectional areas vary inversely with the local magnetic flux density $B\equiv|\bf{B}|$.
Along these tubes, the plasma flow is subject to both internal (pressure gradients) and external (gravity, centrifugal, radiative driving) forces.
Interestingly, in the rigid-field approximation, magnetic and Coriolis forces do not directly influence the dynamics of the flow along field lines since they are always directed perpendicular to the instantaneous velocity vector $\bf{v}$.

\subsection{Euler equations}
\label{sec:Euler}
We can characterize these 1D plasma flows with the conservation form of the quasi-1D Euler equations:
\begin{subequations}
\begin{align}
\label{eq:Eul1}&\frac{\partial\rho}{\partial t} + \frac{1}{A}\frac{\partial}{\partial s}(A\rho v) = 0~,\\
\label{eq:Eul2}&\frac{\partial\rho v}{\partial t} + \frac{1}{A}\frac{\partial}{\partial s}(A\rho v^2) + \frac{\partial P}{\partial s}= \rho(\geff{} + \grad{})~,\\
\label{eq:Eul3}&\frac{\partial\rho \epsilon}{\partial t} + \frac{1}{A}\frac{\partial}{\partial s}(Av(\rho\epsilon+P)) =  \rho v(\geff{}+\grad{}) + \Lambda~,
\end{align}
\end{subequations}
where the independent variables are $t$, the time, and $s$, the arc distance along the field line (relative to an arbitrary zero-point).
The dependent variables are density $\rho$, velocity $v$, pressure $P$, and total energy per unit mass $\epsilon$.
The external sources of energy and momentum are the combined gravitocentrifugal acceleration \geff{} (Section \ref{sec:geff}),  the radiative driving acceleration \grad{} (Section \ref{sec:grad}), and the volumetric energy loss rate $\Lambda\equiv\lamcc{}+\lamrc{}$ due to both radiative cooling (rc) and inverse Compton scattering (cc) (Section \ref{sec:lam_cool}). 

\subsection{Grid geometry}
\label{sec:grid}
The ``arbitrary'' aspect of ARFHD comes from allowing the imposition of any magentic topology, provided that the field lines are physically consistent (i.e. no intersections or discontinuities).
This is an improvement over the original RFHD, which allowed only a dipole topology.

In the reference Cartesian grid comprising the magnetosphere, we define $(0,0,0)$ as the center of the star and the $z$-axis as the stellar rotation pole.
Each field line is a three-dimensional space curve $\mathbf{r}(s)$ parameterized by the arc distance $s$, chosen so that the tangent vector $\hat{\mathbf{s}} = d\mathbf{r}/ds$ is everywhere parallel to the local magnetic field vector $\mathbf{B}$. 
We use the sign of the velocity to indicate the direction of flow; positive (negative) means that the plasma is flowing in the direction of increasing (decreasing) $s$.

The creation of magnetic topologies is outside the scope of this paper, though there has recently been great success in reconstructing magnetic fields of OB stars using surface spectropolarimetry and source-surface reconstruction \citep{donati06, kochukhov11}.

\subsection{Equations of state and energy}
\label{sec:state}
In ARFHD, we assume an ideal gas:
\begin{align}
\label{eq:ideal}P = \frac{\rho k_b T}{\bar\mu}
\end{align}
with the Boltzmann constant $k_b$ and $\bar\mu\equiv\mu u_\mathrm{atm}$ with $u_\mathrm{atm}$ the atomic mass unit.
The mean molecular weight $\mu$ is determined by an expression appropriate to a fully ionized mixture:
\begin{align}
\mu = \left[2X_H + \frac{3}{4}(1-X_H-Z) + \frac{Z}{2}\right]^{-1}
\end{align}
with $X_H$ and $Z$ the usual hydrogen and metal mass fractions. 
Similarly, for a fully ionized plasma, we define a mean molecular weight per hydrogen atom
\begin{align}
\bar\mu_p = u_\mathrm{atm}/X_H,
\end{align}
and a mean molecular weight per free electron
\begin{align}
\bar\mu_e= 2u_\mathrm{atm}/(1+X_H).
\end{align}
The electron scattering opacity is
\begin{align}
\kappa_e = \sigma_T/\bar\mu_e,
\end{align}
with $\sigma_T$ the Thomson scattering cross-section.

The accompanying equation for the total energy per unit mass is:
\begin{align}
\label{eq:epsilon}\epsilon =  \frac{v^2}{2} + \frac{P}{\rho(\gamma-1)}~,
\end{align}
with $\gamma$ the usual ratio of specific heats, $5/3$ for a monatomic gas.

\subsection{Stellar surface properties}
Due to rotation, the stellar surface is centrifugally distorted.
In the Roche approximation, with the assumptions of a point-like mass distribution and uniform rotation, the surface is an equipotential whose radius $R_*$ varies with rotational colatitude $\theta$ as:
\begin{align}
\label{eq:R_surf}\frac{R_*}{R_p} = \frac{3}{\omega\sin\theta}\cos\left[\frac{\pi+\cos^{-1}(\omega\sin\theta)}{3}\right]~.
\end{align}
Here, after defining $\Omega$ as the angular rotation frequency, $M_*$ the stellar mass, and $R_p$ the stellar polar radius,
\begin{align}
\omega \equiv \Omega\sqrt{\frac{27 R_p^3}{8GM_*(1-\gammael{})}}
\end{align}
is the normalized rotation angular frequency, with $\omega = 1$ corresponding to critical rotation.
\gammael{} is the Eddington parameter, defined in the next section.

\subsection{Gravitocentrifugal acceleration}
\label{sec:geff}
The effective gravity, \geff{}, is calculated as the combined gravitocentrifugal acceleration as derived from a scalar effective potential $\Phi_\mathrm{eff}$ and projected along the field line.
The effective gravity vector is
\begin{align}
\label{eq:geff1}\bf{g_\mathrm{eff}} = -\nabla\Phi_\mathrm{eff}~.
\end{align}
Within the Roche approximation, this effective potential is given by
\begin{align}
\Phi_\mathrm{eff} = -(1-\gammael{})\frac{GM_*}{r} - \frac{1}{2}\Omega^2\bar r^2~,
\end{align}
where we take into account the effective reduction in gravity due to the outward force from the electron scattering continuum through the Eddington parameter $\gammael{}\equiv \kappa_eL_*/(4\pi cGM_*)$.
In the centrifugal force term, $\bar{r} = |\bar{\bf{r}}|$ with $\bar{\bf{r}} = [x,y,0]$ the vector drawn from the rotation axis to the position at $\bf{r}$.

In order to obtain \geff{} in \autoref{eq:Eul2}, we need to translate $\bf{g_\mathrm{eff}}$ into an acceleration \textit{along} the field line:
\begin{align}
\geff{} = \mathbf{g_\mathrm{eff}}\cdot\hat{\mathbf{s}} &= -(1-\gammael{})GM_*\frac{\psi}{r^2} + \Omega^2\bar r\bar\psi~\nonumber\\
\label{eq:geff}&= (1-\gammael{})GM_*\left(\frac{-\psi}{r^2} + \frac{8\omega^2}{27 R_p^3}\bar r\bar\psi\right),
\end{align}
with $\psi\equiv \hat{\mathbf{r}} \cdot \hat{\mathbf{s}}$ and likewise $\bar{\psi} \equiv \hat{\bar{\mathbf{r}}} \cdot \hat{\mathbf{s}}$. 
Here, $\hat{\mathbf{r}}$ is the unit radial vector, and $\hat{\bar{\mathbf{r}}}$ is the unit vector parallel to $\bar{\mathbf{r}}$. 

In our rotation analysis (Section \ref{sec:mflux_rot}), we do not take into account the effect of rotational gravity darkening on stellar luminosity \citep{gayley00}.
This will be deferred to future studies.

\subsection{Radiative driving}
\label{sec:grad}
The chief mechanism for wind acceleration is radiation line-driving.
To quantify this, we implement the \citet{owocki88} version of the usual CAK formalism for line-driven stellar winds.
Assuming that the star is a point source of radiation, the acceleration is:
\begin{align}
\label{eq:grad_sat0}\mathbf{g_\mathrm{rad}} = \frac{\kappa_e \bar Q L_*}{4\pi r^2c}\frac{(1+\tau_\mathrm{sob})^{1-\alpha}-1}{(1-\alpha)\tau_\mathrm{sob}}~\hat{\mathbf{r}}~,
\end{align}
where $\bar Q$ is the dimensionless line strength parameter introduced by \citet{gayley95}, $\alpha$ is the CAK-power law index, and
\begin{align}
\label{eq:tau_sob}\tau_\mathrm{sob} \equiv \frac{c\rho\kappa_e\bar Q}{|\delta_v|}
\end{align}
is the Sobolev optical depth.

For $\delta_v$, the local velocity gradient, we follow the same procedure as T07 (see their Section 2.5) and assume that the polar velocity derivative vanishes.
Thus, we adopt the approximation $\delta_v \approx \partial v/\partial s$.

At low $\tau_\mathrm{sob}$, \autoref{eq:grad_sat0} correctly reduces to the optically-thin line force.
This is an improvement over the previous RFHD implementation, which led to an infinitely large radiative acceleration at zero density (see T07 Equation 25).
Finally, we take $\grad{} = \bf{g_\mathrm{rad}}\cdot\hat B$ to get the radiative acceleration along the field line, giving us a final expression:
\begin{align}
\label{eq:grad_sat}\grad{} = \frac{\kappa_e \bar Q L_*}{4\pi r^2c}\frac{(1+\tau_\mathrm{sob})^{1-\alpha}-1}{(1-\alpha)\tau_\mathrm{sob}}~\psi~.
\end{align}

This is a rather simplistic view of line-driven winds, but we emphasize that we are not making any unique insights into the inherent nature of line-driven acceleration.
Rather, we are taking the first steps into understanding how a magnetic field affects a line-driven wind.
For more detailed massive-star wind models, see e.g. \citet{kudritzki02} (modified CAK); \citet{mueller08} (Monte Carlo technique); \citet{sundqvist15} (scattering).

\subsection{Cooling}
\label{sec:lam_cool}
The volumetric cooling rate $\Lambda$ is evaluated as the sum of an inverse Compton cooling term $\lamcc{}$ and a radiative cooling term $\lamrc{}$.
We calculate $\lamcc{}$ from the electron pressure $n_ek_bT$ as per Equation 4 of \citet{white95}:
\begin{align}
\lamcc{} = \frac{-4\sigma_T}{m_ec}n_e k_b T U_\mathrm{rad}
\end{align}
with $U_\mathrm{rad} = L_*/(4\pi r^2 c)$ the stellar radiation energy density and $n_e$ the electron number density.
We calculate \lamrc{}~as:
\begin{align}
\lamrc{} = -n_e n_p\Lambda_\mathrm{rad} = -\frac{\rho^2\Lambda_\mathrm{rad}}{\bar\mu_e\bar\mu_p} = -\rho^2 \Lambda_m
\end{align}
where $n_p$ is the proton number density. 
$\Lambda_\mathrm{rad}$ is the optically thin cooling function, typically obtained in tabular form from a plasma emission code \citep{schure09}, and we define a mass-weighed form $\Lambda_m \equiv \Lambda/\bar\mu_e\bar\mu_p$.

\section{Steady-state wind analysis}
\label{sec:topo}
\citet{townsend07}, with their RFHD approach, simulated colliding wind flows which create reverse shocks that propagate from the apex towards the footprints of each field line.
Within each line, these shocks separate a wind-driving region from a post-shock cooling region which may also contain a centrifugally-supported disk.
The overall field line structure can be considered ``quasi-steady'': the wind-acceleration and cooling regions each reach a steady state, but the shock location oscillates.
In this paper, we analyze only the wind-driving region, which contains the ``critical point'' that sets the steady-state mass-flux and the overall mass-loss rate.
This will also allow us to understand how the magnetic field changes the overall velocity structure of the magnetosphere.

Before we analyze the ARFHD equations presented above (Section \ref{sec:Euler}), we first simplify using several assumptions relevant to the wind-driving region close to the star.
We assume the wind is isothermal, and, following \citet{drew89}, set the temperature $T$ equal to the stellar effective temperature $T_\mathrm{eff}$. 
Also, we assume that the wind remains optically thick ($\tau_\mathrm{sob} \gg 1$) and that it has reached a steady state.

In deriving the magnetospheric wind equation of motion, it is convenient to use the primitive variable form of \autoref{eq:Eul1}-\autoref{eq:Eul3}.
Under our stated assumptions, these equations reduce to
\begin{align}
\label{eq:simp1}\frac{\partial}{\partial s}(A\rho v) &= 0,\\
\label{eq:simp2}v\frac{\partial v}{\partial s} + \frac{1}{\rho}\frac{\partial P}{\partial s} &= \rho g,\\
\label{eq:simp3}P = c_\mathrm{s}^2\rho, 
\end{align}
where $\lambda = \partial A/\partial s/A$ is the areal gradient term and $c_\mathrm{s}^2 = P/\rho = k_b T_\mathrm{eff}/\bar{\mu}$ is the isothermal sound speed.
We can then derive an equation of motion:
\begin{align}
\label{eq:eom}vv'\left(1 - c_\mathrm{s}^2/v^2\right) - \geff{} -c_\mathrm{s}^2\lambda - \grad{} = 0,
\end{align}
where we define $v'\equiv \partial v/\partial s$.

In the optically thick limit, $\grad{}$ reduces to
\begin{align}
\label{eq:grad_thick}\grad{} = \frac{\kappa_{e} \bar{Q} L_{*}}{4\pi r^2 c}\frac{\tau_\mathrm{sob}^{-\alpha}}{1-\alpha} \psi,
\end{align}
which is equivalent to Equation 25 of T07. 
Substituting in the expression (\ref{eq:tau_sob}) for $\tau_\mathrm{sob}$, and then eliminating the explicit dependence on density via the continuity equation (\ref{eq:simp1}), we obtain after some algebra
\begin{align}
\grad{} = \Delta \left(\frac{A}{A_*}\right)^{\alpha} \frac{\psi}{r^{2}} |v v'|^{\alpha}.
\end{align}
Here,
\begin{align}
\Delta &\equiv \frac{\left(\bar Q\gammael{}GM_*\right)^{1-\alpha}}{1-\alpha}\left(\frac{L_*}{4\pi\dot m_* c^2}\right)^\alpha.
\end{align}
parameterizes the mass-loss rate, with $A_{*}$ the area of the flux tube at the stellar surface, and $\dot{m}_{*}$ is the mass flux into the tube.

Now that we have derived a equation of motion (\ref{eq:eom}), we can solve for the values at the critical point.
For simplicity, we shall neglect the Parker term $c_\mathrm{s}^2\lambda$ since it is typically of order $c_\mathrm{s}^2/v_\mathrm{esc}^2\approx 0.001$ relative to the gravitational acceleration term, where $v_\mathrm{esc} = [2GM_*(1-\gammael{})/R_*]^{1/2}$ is the escape velocity at the stellar surface.
Defining
\begin{align}
\label{eq:mot_unsat_simp}F[s,y,u] \equiv y(1-1/u^2) - \geff{} - \Delta\left(\frac{A}{A_*}\right)^\alpha\frac{\psi}{r^2}|y|^\alpha,
\end{align}
where $u \equiv v/c_{\rm s}$ and $y \equiv v v'$, the equation of motion can be written as
\begin{align}
\label{eq:F}F[s, y, u] = 0.
\end{align}

Following CAK and \citet{abbott80}, we fix the wind critical point by the singularity condition
\begin{align}
\label{eq:sing}\frac{\partial F}{\partial y} = 0.
\end{align}
In order that the velocity gradient ${\rm d}v/{\rm d}s$ remain bounded at the critical point, it is also necessary that the regularity condition
\begin{align} 
\label{eq:reg}\frac{\partial F}{\partial s} + \left( \frac{y}{c_{\rm s}^{2} u} \right) \frac{\partial F}{\partial u} = 0
\end{align}
be satisfied at the critical point (this can be derived by taking the total derivative of \autoref{eq:F}).

For a given choice of the parameter $\Delta$, Equations~(\ref{eq:F}--\ref{eq:reg}) can in principle be solved to find the unknowns $(s,y,u)$ at the critical point (if solutions exist).
However, in his analysis of spherical wind outflows, \citet{bjorkman95} found that the critical point location was quite sensitive to the surface mass-flux.
He concluded that it was much easier to fix the location and then solve for the mass flux.
Following his lead, we therefore treat the critical point location $s_{\rm crit}$ as a free parameter, and solve for $\Delta$, $y$ and $s$ at $s = s_{\rm crit}$.
We defer until later (Section \ref{sec:crit}) the question of how to choose $s_{\rm crit}$ appropriately. 

We leave the mathematical derivation of the critical values to Appendix \ref{app:crit_unsat}.
In this derivation, we obtain a special function
\begin{align}
\label{eq:gen_phi_unsat}\frac{y_c}{u_c^2} = \Phi \equiv \pm\sqrt{\frac{\partial \geff{}}{\partial s} - \frac{\geff{}}{1-\alpha}\left[\alpha\lambda + \frac{1}{\psi}\frac{\partial\psi}{\partial s} - \frac{2\psi}{r}\right]}.
\end{align}
Due to our sign convention (Section \ref{sec:grid}), $\Phi$ can be positive or negative, corresponding to a positive/negative $\psi$.
This is because plasma accelerating away from the stellar surface flows in the direction of increasing (decreasing) $s$ for positive (negative) $\psi$.

Either way, we solve for the critical velocity:
\begin{align}
\label{eq:gen_unsat_u}u_c^2 = 1 \mp\frac{\sqrt{2}\alpha}{(1-\alpha)c_\mathrm{s}\Phi}\geff{},
\end{align}
with the top (bottom) resulting from the positive (negative) root of \autoref{eq:gen_phi_unsat}.
Similarly, we obtain the critical $y$:
\begin{align}
\label{eq:gen_unsat_y}y_c = \pm\frac{c_\mathrm{s}\Phi}{\sqrt{2}} - \frac{\alpha}{1-\alpha}\geff{}.
\end{align}
Remembering that our $y = vv'$, this critical value is nearly identical to CAK Equation 45 in the zero sound-speed limit ($\Phi\to 0$), with differences coming from a factor $\psi$ resulting from non-radial paths and from rotational acceleration.

Finally, we can solve for our critical eigenvalue and surface mass-flux:
\begin{align}
\label{eq:gen_unsat_Delta}\Delta_c &= -\frac{\geff{}}{(1-\alpha)}\left(\frac{A_*}{A}\right)^\alpha\frac{r^2}{\psi|y_c|^\alpha},\\
\label{eq:gen_unsat_mflux}\dot m_* &= \left(\frac{(\bar Q\gammael{}GM_*)^{1-\alpha}}{1-\alpha}\right)^{1/\alpha}\frac{L_*}{4\pi c^2}\frac{1}{\Delta_c^{1/\alpha}},
\end{align}
A similar procedure for the general critical point values is presented in Appendix \ref{app:crit_sat}.

As noted by \citet{marlborough84} and \citet{cure04}, this type of analysis implies a range to the allowable critical point position.
Here, we require that $\Phi^2 > 0$ and $\Delta_c >0$ at \scrit{} for a trans-critical steady-state wind.

\begin{table*}
\centering
\caption{\label{tab:star}Stellar and wind parameters used throughout this paper to represent a typical magnetic B-type star with a centrifugal magnetosphere and an O-type star with a dynamical magnetosphere. Values are taken from Table 2 in \citet{bjorkman95} for the B-star and \citet{uddoula14} for the O-star.}
\begin{tabular}{ccccccccc}
\hline\hline
Type & $M_*$& $R_*$ & $T_\mathrm{eff}$ & $\alpha$ & \gammael{} & $\bar Q$ & $B_p$ & $\eta_*$\\\hline
$B$&$9.0~M_\odot$ & $4.5~R_\odot$ & $21000~\mathrm{K}$ & 0.56 & 9.27\e{-3} & 1025.14 & 11 kG & 4.29\e{5}\\
\hline
$O$&$50~M_\odot$&$19~R_\odot$&$41860~\mathrm{K}$& 0.6 & 0.5 & 500 & 3.715 kG &100
\end{tabular}
\end{table*}

Although our ansatz assumes a magnetic star, we can pretend there is no field by defining a radial magnetic topology with a spherically expanding cross-section.
This results in $\psi = 1$ and $A_*/A = R_*^2/r^2$.
In the limit of zero rotation and zero sound speed ($\Phi \to 0$), making these subsitutions in \autoref{eq:gen_unsat_Delta} yields
\begin{align}
\Delta_c &= \frac{(1-\gammael{})GM_*}{(1-\alpha)r^2}\frac{R_*^{2\alpha}}{r^{2\alpha-2}}\frac{|r^{2\alpha}|}{|\frac{\alpha}{1-\alpha}(1-\gammael{})GM_*|^\alpha}\nonumber\\
&= \frac{1}{[\alpha/(1-\alpha)]^\alpha}\frac{[GM_*(1-\gammael{})]^{1-\alpha}R_*^{2\alpha}}{1-\alpha},
\end{align}
and a surface mass-flux
\begin{align}
\dot m_* &= \frac{\alpha}{1-\alpha}\left(\frac{\bar Q\gammael{}}{1-\gammael{}}\right)^{\frac{1-\alpha}{\alpha}}\frac{L_*}{4\pi R_*^2c^2}.
\end{align}
As defined by \citet{gayley95} using his $\bar Q$ formalism, the CAK mass-loss rate is
\begin{align}
\Mcak{}\equiv \frac{\alpha}{1-\alpha}\left(\frac{\bar Q\gammael{}}{1-\gammael{}}\right)^{\frac{1-\alpha}{\alpha}}\frac{L_*}{c^2},
\end{align}
and we see that our derived surface mass-flux is $\dot m_* = \Mcak{}/(4\pi R_*^2)$.
This demonstrates that our general equations correctly reproduce the usual CAK mass-loss rate in the proper limit.

\section{Critical point calculation}
\label{sec:crit}
The critical point location, \scrit{}, is required to accurately calculate the surface mass-flux, which, through the density, sets the level of radiative driving and emission throughout the magnetosphere.
Since \scrit{} is a free parameter in the above critical point calculation, we must provide a boundary condition to obtain \scrit{} and the resulting critical surface mass-flux.
This is especially important for calculating a dipole star's mass-loss rate, since the critical mass-flux is sensitive to the critical location (Section \ref{sec:mflux_zero}).
Following Equation 24 of \citet{bjorkman95}, we define the boundary density such that the resultant electron scattering optical depth $\tau_{es} \approx 1$ at the stellar surface.
Thus,
\begin{align}
\label{eq:boundary_rho}\rho_0 = \frac{\tau_{es}}{\kappa_e H} \approx \frac{(1-\gammael{})GM_*}{\kappa_e R_*^2 c_\mathrm{s}^2}
\end{align}
where $H = c_\mathrm{s}^2/g = c_\mathrm{s}^2 R_*^2/[(1-\gammael{})GM_*]$ is the pressure scale height, corrected for the electron scattering acceleration.
Using the continuity equation and the critical surface mass flux, we can solve for the boundary velocity:
\begin{align}
\label{eq:boundary_v}v_0 = \frac{\rho_* v_*}{\rho_0} = \frac{\dot m_*}{\rho_0}
\end{align}
where $\dot m_*\equiv\rho_* v_*$ is defined as the surface mass-flux into the field line.
Since this introduces a dependence on \scrit{} for both the boundary and critical point velocities, we must use an iterative algorithm to satisfy both conditions simultaneously.

The set of equations we use for the integration are the wind equation of motion (\autoref{eq:eom}), the steady-state continuity equation $d\rho/ds = -\rho[\lambda + (dv/ds)/v]$, and the isothermal approximation $dP/ds = c_\mathrm{s}^2 d\rho/ds$.
\autoref{eq:eom} is not easily solved for $v'$, however, since there are multiple roots.
There are usually three roots: two positive and one negative.
This differs slightly from \citet{kudritzki02}, who found two roots; we find an extra one since $\tau_\mathrm{sob}$ has a dependency on the absolute value of $v'$.
Inside of the sonic point, though, there are instead one positive and two negative roots.
We always choose a positive root in order to enforce an accelerating outflow.
When multiple positive roots exist, we choose the smaller root in the subcritical region and the larger root beyond the critical point \citep{cassinelli79, abbott80}.

Our iterative algorithm for calculating \scrit{} is as follows:
\begin{enumerate}
\item Choose trial critical point location.
\item Calculate velocity, mass-flux, density at the critical point.
\item Integrate to boundary, taking the smallest positive root of the possible velocity derivatives.
\item If resulting boundary velocity is too high, move \scrit{} out. Else if too low, move \scrit{} in.
\item Repeat from step 2 until correct boundary values are reached.
\end{enumerate}

As a check, we calculate the critical radius and mass-flux for a straight, spherically-diverging flux tube ($\psi = 1$; $A_*/A = R_*^2/r^2$) for the stellar parameters chosen in \citet{bjorkman95} (\autoref{tab:star}).
We calculate $\rcrit{}= \scrit{}=1.5589 R_*$, which matches well with \citet{bjorkman95}'s derived value of $1.5594$.
The resulting mass-flux, $\dot m_* = 9.2516\e{-8} \unit{g/cm^2}$ also fits with his derived surface mass-flux $\dot M/4\pi R_p^2 = 9.249\e{-8}\unit{g/cm^2}$.

\section{Magnetic dipole model}
\label{sec:dip_topo}
Now that we have developed our general critical equations, we now derive the critical values for a wind channeled by an magnetic dipole whose pole is aligned with the rotation axis.
This field forces the plasma to co-rotate with the star (i.e. a magnetosphere).
Instead of assuming a radial outflow, we force the plasma to flow along the magnetic flux tubes.
Additionally, we define the combined gravitocentrifugal force in the same manner as the ARFHD formulation (Section \ref{sec:geff}), i.e. with rigid-body rotation.

While it is possible to solve the critical point equations with $s$ as the independent variable, it is relatively more convenient here to parameterize the spatial variables with $\tilde\theta$, the magnetic colatitude.
We do this because although the plasma flows along the magnetic field line, most of our external forces are dependent on $r$.
It is easier to set $\tilde\theta$ as the spatial variable rather than have to solve for $r$ in terms of $s$. 

In an aligned dipole, the magnetic pole is parallel to the rotational pole (the $z$-axis in our coordinate system; Section \ref{sec:grid}), so we take $\tilde\theta = \theta$, where $\theta$ is the rotational colatitude.
First, we start with the definition of an aligned dipole field (e.g. T07):
\begin{subequations}
\begin{align}
\label{eq:dipole_B}\bf{B} &= \frac{B_0}{2(r/R_p)^3} \left(2\cos\theta \hat r + \sin\theta \hat \theta\right),\\
\label{eq:dipole_Bhat}\hat B &= \frac{\bf{B}}{|\bf{B}|} = \frac{2\cos\theta\hat r + \sin\theta\hat \theta}{\sqrt{1+3\cos^2\theta}}.
\end{align}
\end{subequations}
We note that $R_p$ is the polar radius of the star, not the stellar surface radius $R_*$ (\autoref{eq:R_surf}).
From the parametric equation of a dipole field line (e.g. \citealp{babel97}), we have
\begin{align}
\label{eq:r_theta}r(\theta) = R_p L\sin^2\theta = r_m\sin^2\theta, 
\end{align}
where $r_m\equiv R_p L$ is the maximum extent of the field line and $L$ is the magnetic shell parameter.
Each individual line exists over the range $\theta_m < \theta < \pi - \theta_m$, with $\theta_m = \sin^{-1}\sqrt{R_*/(R_pL)}$ marking the northern magnetic footprint and $\pi - \theta_m$ marking the southern.
Each field line can be uniquely identified by $L$ and its magnetic azimuthal coordinate which denotes the half-plane containing that line.
For our aligned dipole model, we will place every individual line in the same half-plane and assume azimuthal symmetry.
Thus, knowing $L$ or $\theta_m$ is sufficient for identifying a particular line.

We can obtain the path length $s$ along the line with
\begin{align}
\label{eq:ds_theta}ds^2 = dr^2 + r^2 d\theta^2 = r_m^2\sin^2\theta(1+3\cos^2\theta)d\theta^2,
\end{align}
which, after integrating, yields
\begin{align}
s = -\frac{r_m}{2}\left[\frac{\sinh^{-1}(\sqrt{3}\cos{\theta})}{\sqrt{3}} + \cos{\theta}\sqrt{1+3\cos^2{\theta}}\right] + \mathrm{const}.
\end{align}
We select our constant of integration to enforce $s=0$ where the magnetic field comes out of the stellar surface: the northern footprint ($\theta=\theta_m$).

With these definitions of $r$ and $s$, we can write all of the spatial variables as functions of $\theta$ (Appendix \ref{app:dip}).
These can be then be used to solve the critical point values derived in Section \ref{sec:topo}.
The general critical point values (Appendix \ref{app:crit_sat}) can also be parameterized in this manner, using the same spatial variables.

With our general critical point algorithm established, we now turn our attention to how an aligned dipole magnetic field affects the stellar wind, namely its mass-loss rate and terminal velocity.
Additionally, we will study how the stellar rotation rate influences the CAK critial point and resulting wind properties.
For this analysis, we generate a grid of 500 dipole field lines with footprints covering the northern hemisphere of the star ($0 < \theta < \pi/2$) in linear space.
This is repeated for several critical rotation fractions $\omega=[0.0,0.2,0.35,0.5,0.65,0.8]$, giving a total of 3000 dipole lines.
Field lines with $L > 100 R_p$ are arbitrarily truncated at $R = 100 R_p$; the rest of the lines extend to the magnetic equator.
This truncation does not affect the critical value calculations since that only depends on the boundary condition and the field line geometry inside of the critical point.
Additionally, as we later show, this trunctation radius is larger than the ``closure radius'' of our model magnetospheres, so our apex velocity calculations will not be affected.

We calculate the critical point location and resulting surface mass-fluxes using the procedure described in Section \ref{sec:crit}. 
After finding the critical values, we integrate from the critical point to the apex of the field line to obtain the apex velocity, $v_\mathrm{apex}$.
We do this for two different sets of stellar parameters, one representing a centrifugal magnetosphere and one a dynamical magnetosphere (\autoref{tab:star}).
For the centrifugal magnetosphere star, we use parameters from \citet{bjorkman95} representing an early-B star similar to the archetype $\sigma$ Orionis E.

For the other star, we follow the MHD simulations of \citet{uddoula02} and choose a $\zeta$ Puppis analogue, representing a dynamical magnetosphere: $M_\mathrm{eff} = 25~M_\odot,~R=19~R_\odot,~T_\mathrm{eff}=41860~\unit{K}$ (such that $L_\mathrm{star} = 10^6 L_\odot$), $\alpha = 0.6$, $\bar Q = 500$, and we take solar values for the mass fractions $X_H$ and $Z$.
Since the stellar mass above is an effective mass and already takes into account the factor of two reduction below the Newtonian mass due to the electron scattering continuum force, we take $M = 50 M_\odot$ and $\gammael{} = 0.5$ in our model.

Since our ansatz assumes an infinite magnetic confinement, the actual magnitude of the dipole field (i.e. $B_0$ in \autoref{eq:dipole_B}) only matters when estimating which lines are in the closed magnetosphere (\autoref{eq:r_close}).

\section{Surface mass flux}
\label{sec:mass_flux}
\subsection{Zero rotation in the optically-thick limit}
\label{sec:mflux_zero}
Since magnetic dipole field lines do not come straight out of the stellar surface, the surface mass-flux is tilted relative to a radial mass-flux.
Inspired by the MHD simulations presented in \citet{uddoula02}, \citet{owocki04} (hereafter OD04) used a simple, one-dimensional flow analysis to calculate that the radial mass-flux, $\dot m_r$, scales as
\begin{align}
\label{eq:flux_scaling}\dot m_r = \mu_B\dot m_*=\mu_B^2\mcak{},
\end{align}
where the CAK surface mass-flux is defined as
\begin{align}
\label{eq:mflux_cak}\mcak{} = \frac{\Mcak{}}{4\pi R_p^2} = \frac{L_*}{4\pi R_p^2 c^2}\frac{\alpha}{1-\alpha}\left(\frac{\bar Q \Gamma_{\mathrm{el}}}{1-\Gamma_{\mathrm{el}}}\right)^{(1-\alpha)/\alpha},
\end{align}
and $\mu_B = \hat{\mathbf{n}} \cdot \hat{\mathbf{B}}$ with $\hat{\mathbf{n}}$ the unit vector normal to the stellar surface.
For a non-rotating star, $\hat{\mathbf{n}} = \hat{\mathbf{r}}$.
One factor of $\mu_B$ results from the geometric projection of the tilted flow onto the stellar surface normal ($\dot m_r = \mu_B\dot m_*$).
The other factor results from projecting a radial radiative line force along the field line.
The tension in the magnetic field line negates any acceleration normal to the line, further lowering the critical mass-flux.

We now check this scaling analysis with our dipole model.
For simplicity, we note that $c_s\Phi/\sqrt{2} \ll \alpha/(1-\alpha)~\geff{}$ and take
\begin{align}
\label{eq:y_c_approx}y_c \approx -\frac{\alpha}{1-\alpha}\geff{}.
\end{align}
From \autoref{eq:dipole_A_star_A} we obtain
\begin{align}
\left(\frac{A_*}{A}\right)^\alpha =\left(\frac{R_*}{r_c}\right)^{3\alpha}\left(\sqrt{\frac{1+3\cos^2\theta_{c}}{1+3\cos^2\theta_m}}\right)^\alpha
\end{align}
where $\theta_c$ is evaluated at the critical radius $r_c$ for a given field line and $R_*$ is the stellar radius at the footprint colatitude $\theta_m$.

Combining our eigenvalue relation (\autoref{eq:gen_unsat_Delta}) with the above equations, we obtain
\begin{align}
\Delta_c \approx \frac{-\geff{}|\geff{}|^{-\alpha}}{(1-\alpha)(\frac{\alpha}{1-\alpha})^\alpha}\frac{R_*^{3\alpha} r^{2-3\alpha}_c}{\psi_c}\left(\sqrt{\frac{1+3\cos^2\theta_c}{1+3\cos^2\theta_m}}\right)^\alpha,
\end{align}
where $\psi_c$ is evaluated at the critical point. Finally, our surface mass-flux is
\begin{align}
\label{eq:mflux_approx_unsat}\dot m_* &\approx \frac{\alpha}{1-\alpha}\frac{L_*}{4\pi c^2}|\geff{}|\left[\frac{(\bar Q\gammael{}GM_*)^{1-\alpha}}{-\geff{}}\right]^{1/\alpha}\nonumber\\
&~~~~~\times \frac{\psi_c^{1/\alpha}r^{3-2/\alpha}_c}{R_*^3}\sqrt{\frac{1+3\cos^2\theta_m}{1+3\cos^2\theta_c}}.
\end{align}
For zero rotation, $\geff{} = -(1-\gammael{})GM_*\psi/r^2_c$ and $R_* = R_p$:
\begin{align}
\dot m_* &\approx \frac{\alpha}{1-\alpha}\frac{L_*}{4\pi c^2}\left[\frac{(\bar Q\gammael{}GM_*)^{1-\alpha}}{[(1-\gammael{})GM_*]^{1-\alpha}}\right]^{1/\alpha}\nonumber\\
&~~~~~\times\frac{\psi_c r_c}{R_p^3}\sqrt{\frac{1+3\cos^2\theta_m}{1+3\cos^2\theta_c}}\nonumber\\
\label{eq:mflux_zero_unsat}&\approx\mdip{}\psi_c\sqrt{\frac{1+3\cos^2\theta_m}{1+3\cos^2\theta_c}},
\end{align}
where we will define $\mdip{} = \mdip{}(r_c)$ as the zero-tilt, zero-rotation, optically-thick surface mass-flux for a magnetic dipole:
\begin{align}
\mdip{} &\equiv \frac{\alpha}{1-\alpha}\frac{L_*}{4\pi c^2}\left(\frac{\bar Q\gammael{}}{1-\gammael{}}\right)^{\frac{1-\alpha}{\alpha}}\frac{r_c}{R_p^3}\nonumber\\
\label{eq:mdip}&=\mcak{}\frac{r_c}{R_p}.
\end{align}
We can thus think of \mdip{} as the CAK surface mass-flux corrected for dipole divergence.

We can generalize this straight-line base term for any magnetically-induced areal expansion with
\begin{align}
\dot m_\mathrm{gen} \equiv \mcak{} \left(\frac{r_c}{R_p}\right)^{q-2},
\end{align}
when the areal expansion is proportional to $r^q$, $q=3$ for a dipole.

For our model B star, we calculate $\dot m_\mathrm{dip}\approx 9.53\e{-8}\unit{g/cm^2}$ at the pole, with $r_c = 1.033~R_p$.
If we keep the $\Phi$ term instead of neglecting it (\autoref{eq:gen_unsat_y}), the model calculated mass-flux ($\mtz{}$) is boosted by about 2\%, to $\mtz{} = 9.709\e{-8} \unit{g/cm^2}$.
For our model O star, the polar values are $\mdip{}\approx 1.98\e{-5}\unit{g/cm^2}$ (with $r_{c, theta=0}\approx 1.054~R_p$) and $\mtz{} = 2.05\e{-5}\unit{g/cm^2}$, a difference of about 3\%.
The O-type star has a larger correction than the B-star because of its faster sound speed.

\begin{figure}
\includegraphics[width=\columnwidth]{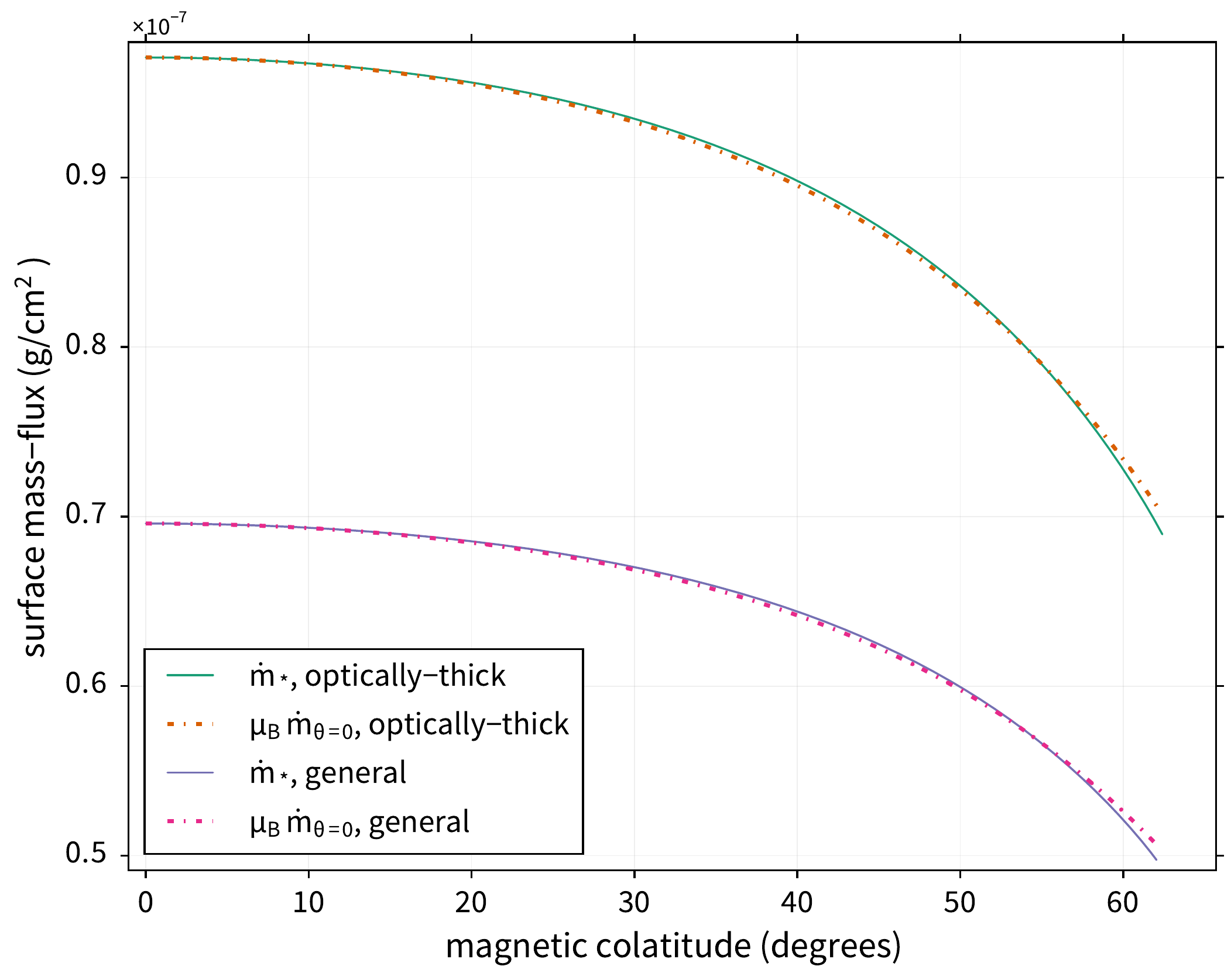}
\caption{\label{fig:M_flux_zero}Surface mass-flux ($\dot m_*$) as a function of line surface colatitude calculated using both the general and optically-thick line-acceleration for a B-type star (\autoref{tab:star}). $\dot m_*$ is compared to a general scaling $\mu_B\mtz{}$ as derived by OD04 (\autoref{eq:flux_scaling}), with a different \mtz{} for both cases. We find an excellent agreement between the scaling and model, with similar results for the O-type star not shown.}
\end{figure}

We can reproduce \autoref{eq:flux_scaling} with several simplifications, which end up canceling each other out.
First, we take $r_c = \rcritz{}$ as constant for every field line (justified in Section \ref{sec:rcrit}).
Next, we assume that the critical radius is very close to the star ($\rcritz{} - R_p \ll R_p$), which allows us to take $\theta_c\approx\theta_m$ and $\psi_c\approx\psi_m = \mu_B$.
Finally, we correct for neglecting the $\Phi$ term by replacing \mdip{} with \mtz{} to obtain the scaling relation:
\begin{align}
\label{eq:flux_scaling_unsat}\dot m_* \sim \mu_B\mtz{}.
\end{align}
Rather conveniently, it turns out that replacing $\theta_c$ and $\psi_c$ with the surface values $\theta_m$ and $\psi_m$ produces opposite effects which nearly cancel each other out.
Overall, we are able to reproduce OD04's general scaling at zero rotation (\autoref{fig:M_flux_zero}), though keeping the exact angular expressions with the constant $r_c$ assumption gives an even better fit.

\subsection{Zero rotation in the general case}
\label{sec:mflux_sat_zero}
In the general line-force critical analysis (Appendix \ref{app:crit_sat}), an important parameter arises:
\begin{align}
\chi\equiv\left(1+\tau_\mathrm{sob}\right)^{-\alpha}.
\end{align}
Physically, it represents the ``transparency'' of the wind where low $\chi$ means a optically thick wind and high $\chi$ means a optically thin wind.
Alternatively, it represents a ``optically-thin correction level'' parameter which characterizes the relative importance of using the general form of \grad{} rather than its optically-thick limit (\autoref{eq:grad_thick}).
For optically-thick winds, $\tau_\mathrm{sob} \gg 1$ and the line-acceleration reduces to $\gradun{}\propto\tau_\mathrm{sob}^{-\alpha}$.
For optically-thin winds, $\tau_\mathrm{sob} \to 0$ and $\chi\to 1$.

We can repeat the previous section's scaling analysis for a general line force (derivation in Appendix \ref{app:mflux_scaling_sat}).
With the same approximations as above and zero rotation, we get the scaling
\begin{align}
\label{eq:mdip_sat}\dot m_*\approx\mdip{}(r_c)\sigsatz{}\psi_c\sqrt{\frac{1+3\cos^2\theta_m}{1+3\cos^2\theta_c}},
\end{align}
where we have defined an optically-thin-correction (OTC) parameter
\begin{align}
\label{eq:sat_param}\sigsatz{} \equiv \mdipsat{}(r_c)/\mdip{}(r_c) = \frac{\left|1-\alpha-\left[\frac{1-\chi_0^{1/\alpha-1}}{1-\chi_0^{1/\alpha}}\right]\right|}{\alpha},
\end{align}
with $\chi_0$ as the value of $\chi$ evaluated at the critical point for zero rotation.
Notably, the critical value of $\chi$ is set by the ratio between the non-radiative external forces (gravity, centrifugal, areal gradient) and the optically thin radiative force (c.f. \autoref{eq:step_sat3}).
With zero rotation, $\chi_0=(1-\gammael{})/(\gammael{}\bar Q)$ is independent of surface colatitude, which allows us to use a constant \sigsatz{} across the stellar surface.

Essentially, \sigsatz{} results from the error in assuming an optically-thick wind.
In O stars, the increased luminosity drives a much higher surface mass-flux, leading to a more optically-thick wind than in B stars.
Thus, $\chi_0$ is smaller and \sigsatz{} is closer to unity for more massive stars.

\sigsatz{} allows us to correct our mass-flux estimates, though the critical radius (and thus the base mass-flux) will be different between the general and optically-thick cases (Section \ref{sec:rcrit}).
For an optically thick wind, $\chi \to 0$ and $\sigsatz \to 1$, reproducing \mdip{}.
For an optically thin wind, using l'H\^opital's rule yields
\begin{align}
\lim\limits_{\chi\to1} \sigsatz = \frac{|1-\alpha-(1-\alpha)|}{\alpha} = 0,
\end{align}
which is expected since the optically-thick line force goes to infinity as the density goes to zero.

For our model B star, we calculate $\sigsatz{} \approx 0.725$ and use the polar critical radius $r_{c,\theta=0}\approx 1.0367$ to obtain $\mdip{}\approx 9.56\e{-8}\unit{g/cm^2}$.
The resulting estimated mass-flux, $\sigsatz{}\mdip{}\approx 6.90\e{-8}\unit{g/cm^2}$ compares well with the model-calculated $\mtz{} \approx 6.96\e{-8}\unit{g/cm^2}$.
Similarly, for the O star, we calculate $\sigsatz{}\approx 0.974$, $\mdip{}\approx 1.98\e{-5}\unit{g/cm^2}$ (with $r_{c,\theta=0}\approx 1.054~R_p$), and $\sigsatz{}\mdip{}\approx 1.93\e{-5}\unit{g/cm^2}$.
The model polar mass-flux is $\mtz{}\approx 2\e{-5}\unit{g/cm^2}$.

Simplifying \autoref{eq:mdip_sat} with $\rcritz{} - R_p \ll R_p$, $\theta_c\approx\theta_m$, and $\psi_c\approx\psi_m = \mu_B$ yields the scaling 
\begin{align}
\dot m_* &\approx\mu_B\sigsatz{}\mdip{}(r_c).
\end{align}
We can correct for approximations made in deriving this equation (Appendix \ref{app:mflux_scaling_sat}) by using the model-calculated $\mtzsat{}$ instead of $\sigsatz{}\mdip{}$:
\begin{align}
\label{eq:flux_scaling_sat}\dot m_* \sim\mu_B\mtzsat{}.
\end{align}
We use \mtzsat{} rather than \sigsatz{}\mtz{} because the critical point location is different between the two cases.
As in the optically-thick case, our model magnetosphere shows excellent agreement for this OD04-type scaling (\autoref{fig:M_flux_zero}).

\subsection{With rotation in the optically-thick limit}
\label{sec:mflux_rot}
\begin{figure}
\includegraphics[width=0.49\textwidth]{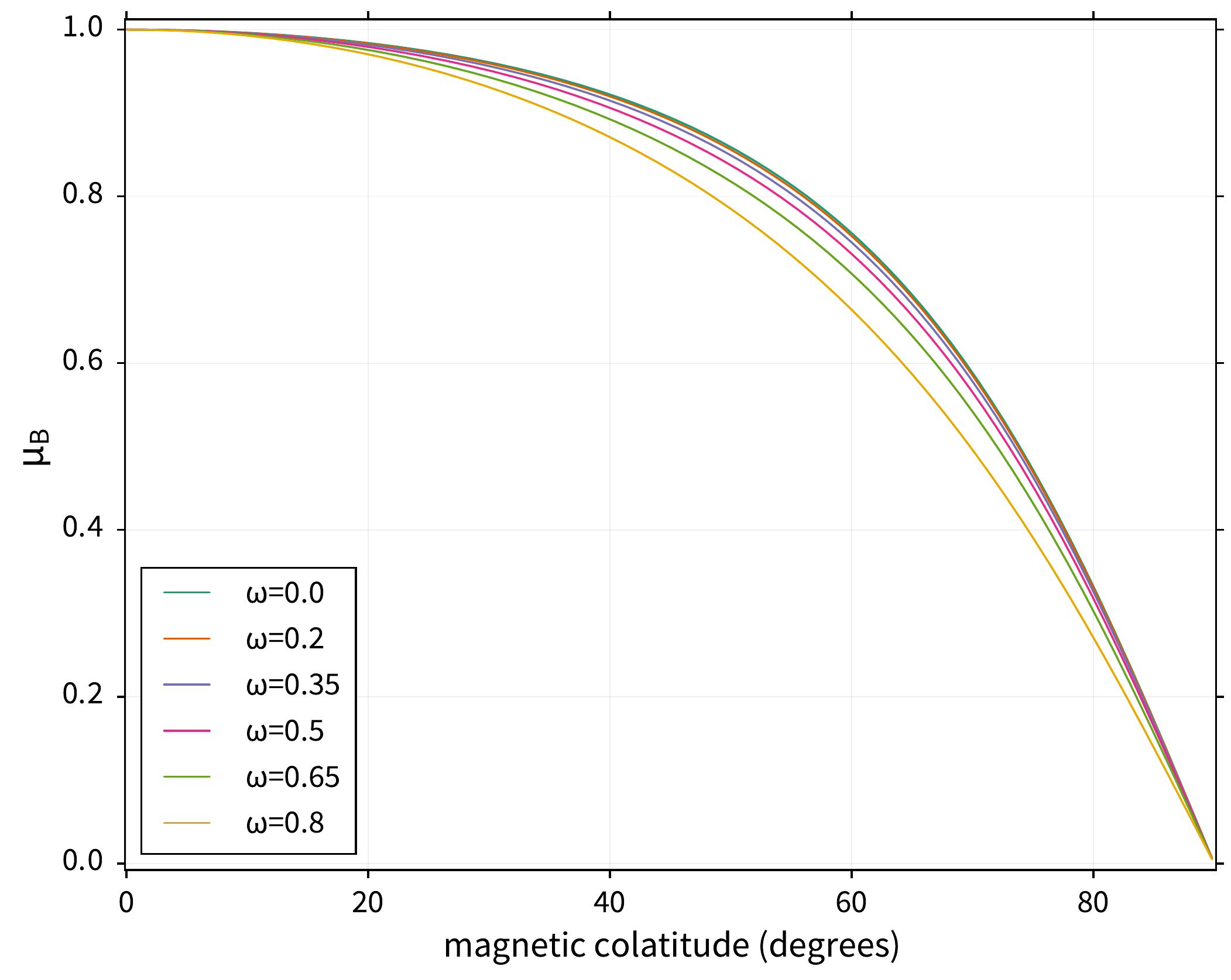}
\caption{\label{fig:mu_normal}Effect of rotation on $\mu_B$, the dot product between the surface normal unit vector and the surface magnetic field unit vector (\autoref{eq:mub}).}
\end{figure}

The previous scaling results, however, depend on zero rotation.
There are two ways rotation changes the above scaling: $\mu_B$ and the location of \rcrit{} (c.f. Section \ref{sec:rcrit}).
For an oblate star, the surface normal unit vector is:
\begin{align}
\label{eq:nhat}\hat n = \frac{\hat r - R'_*/R_*~\hat\theta}{\sqrt{1+\left(R'_*/ R_*\right)^2}}
\end{align}
where $R'_* \equiv \partial R_*/\partial\theta$.
Using \autoref{eq:dipole_Bhat}, the resulting $\mu_B$ is
\begin{align}
\label{eq:mub}\mu_B = \frac{2\cos\theta_m - \sin\theta_m R'_*/R_*}{\sqrt{(1+3\cos^2\theta_m)(1+\left(R'_*/ R_*\right)^2)}},
\end{align}
where, from taking the derivative of \autoref{eq:R_surf},
\begin{align}
\frac{1}{R_p}\frac{\partial R_*}{\partial\theta} &= \frac{\cot\theta\sin\left\{\frac{1}{3}[\pi+\arccos(\omega\sin\theta)]\right\}}{\sqrt{1-\omega^2\sin^2\theta}} \nonumber\\
~~~~&- \frac{3\cot\theta\csc\theta\cos\left\{\frac{1}{3}[\pi+\arccos(\omega\sin\theta)]\right\}}{\omega}.
\end{align}
For a non-rotating star, $R'_* = 0$ and $\mu_B$ is identical to OD04.
$\mu_B$ is affected most at the middle colatitudes, where the stellar surface normal tilts the farthest from the radial direction (\autoref{fig:mu_normal}).
\begin{figure}
\includegraphics[width=0.49\textwidth]{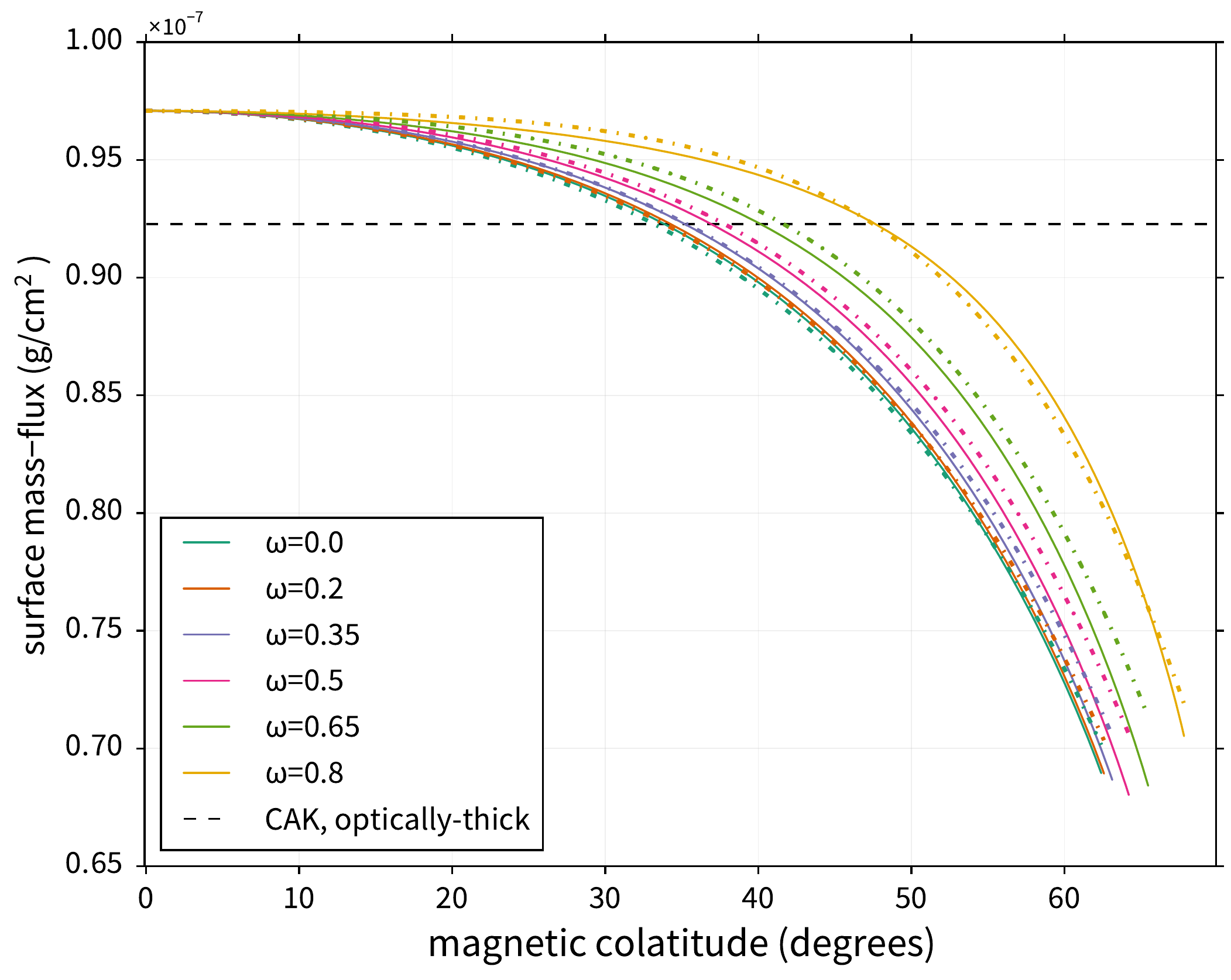}
\caption{\label{fig:M_flux_rot_unsat}Surface mass-flux ($\dot m_*$; solid line) as a function of line surface colatitude for an optically-thick line force and a B-type star (\autoref{tab:star}). $\dot m_*$ is compared to the rotation scaling (dot-dashed line) derived in \autoref{eq:mflux_rot_scaling} and the CAK mass-flux (\autoref{eq:mflux_cak}; dashed line)}
\end{figure}

However, even though $\mu_B$ gets smaller with increased rotation at the middle colatitudes, the surface mass-flux does not decrease in the same manner.
Instead, the rotation of the star boosts the mass-flux above this naive scaling, and, for sufficiently high rotation, actually causes the flux to \textit{increase} as one moves towards the middle colatitudes.
This is chiefly due to the decreased \geff{} as the centrifugal acceleration increases.

We can derive a simple scaling for how $\dot m_*$ depends on rotation, though we will no longer be able to assume a constant \rcrit{}.
We follow the same procedure as in Section \ref{sec:mflux_zero}, but this time we keep the rotation.
First, we rewrite the effective gravity (\autoref{eq:geff}) using our dipole magnetosphere parameterizations on $\theta$ (Appendix \ref{app:dip}), specifically $\bar r = r\sin\theta = R_pL\sin^3\theta$ and $\bar\psi/\psi = 3\sin\theta/2$:
\begin{align}
\geff{} &= -(1-\gammael{})\frac{GM_*\psi}{r^2}\left(1 - \frac{8\omega^2}{27R_p^3}\frac{\bar r r^2\bar\psi}{\psi}\right)\nonumber\\
&=-(1-\gammael{})\frac{GM_*\psi}{r^2}\left(1 - \frac{4}{9}\omega^2L^3\sin^8\theta\right)\nonumber\\
&=-(1-\gammael{})\frac{GM_*\psi}{r^2}\aleph,
\end{align}
where we have defined a rotation effect parameter
\begin{subequations}
\begin{align}
\label{eq:rot_param}\aleph&\equiv 1 - \frac{8\omega^2}{27R_p^3}\frac{\bar r r^2\bar\psi}{\psi}\\
&=1 - \frac{4}{9}\omega^2L^3\sin^8\theta,
\end{align}
\end{subequations}
where the first line is the general definition and the second is specifically for an aligned dipole.
We note that for zero rotation, $L = 1/\sin^2\theta_m$, but the stellar oblateness due to rotation means that this equation no longer applies.
Instead, we combine \autoref{eq:R_surf} and \autoref{eq:r_theta} into
\begin{align}
L = \frac{3}{\omega\sin^3\theta_m}\cos\left[\frac{\pi+\cos^{-1}(\omega\sin\theta_m)}{3}\right],
\end{align}
and get an aligned dipole rotation parameter
\begin{align}
\aleph \equiv 1 - \frac{12\cos^3\left[\frac{1}{3}(\pi+\cos^{-1}(\omega\sin\theta_m))\right]\sin^8\theta}{\omega\sin^9\theta_m}.
\end{align}

The above analysis simply multiplies each instance of \geff{} in \autoref{eq:mflux_approx_unsat} by a factor $\aleph$, and yields a mass-flux estimate
\begin{align}
\label{eq:mflux_rot}\dot m_*\approx\mdip{}(r_c)\left(\frac{R_p}{R_*}\right)^3\aleph_c^{1-1/\alpha} \psi_c\sqrt{\frac{1+3\cos^2\theta_m}{1+3\cos^2\theta_c}},
\end{align}
where $\aleph_c$, $\psi_c$ and $\theta_c$ are evaluated at the critical point and we take into account stellar oblation due to rotation.
Comparing the above approximation against numerical results, we find an error of only $\approx$ 3.5\% (6\%) error in the middle colatitudes for a B-type (O-type) star at $\omega = 0.8$.
This increase in error relative to the non-rotating case comes from neglecting $\Phi$, which is larger at faster rotation rates.

\begin{figure}
\includegraphics[width=0.49\textwidth]{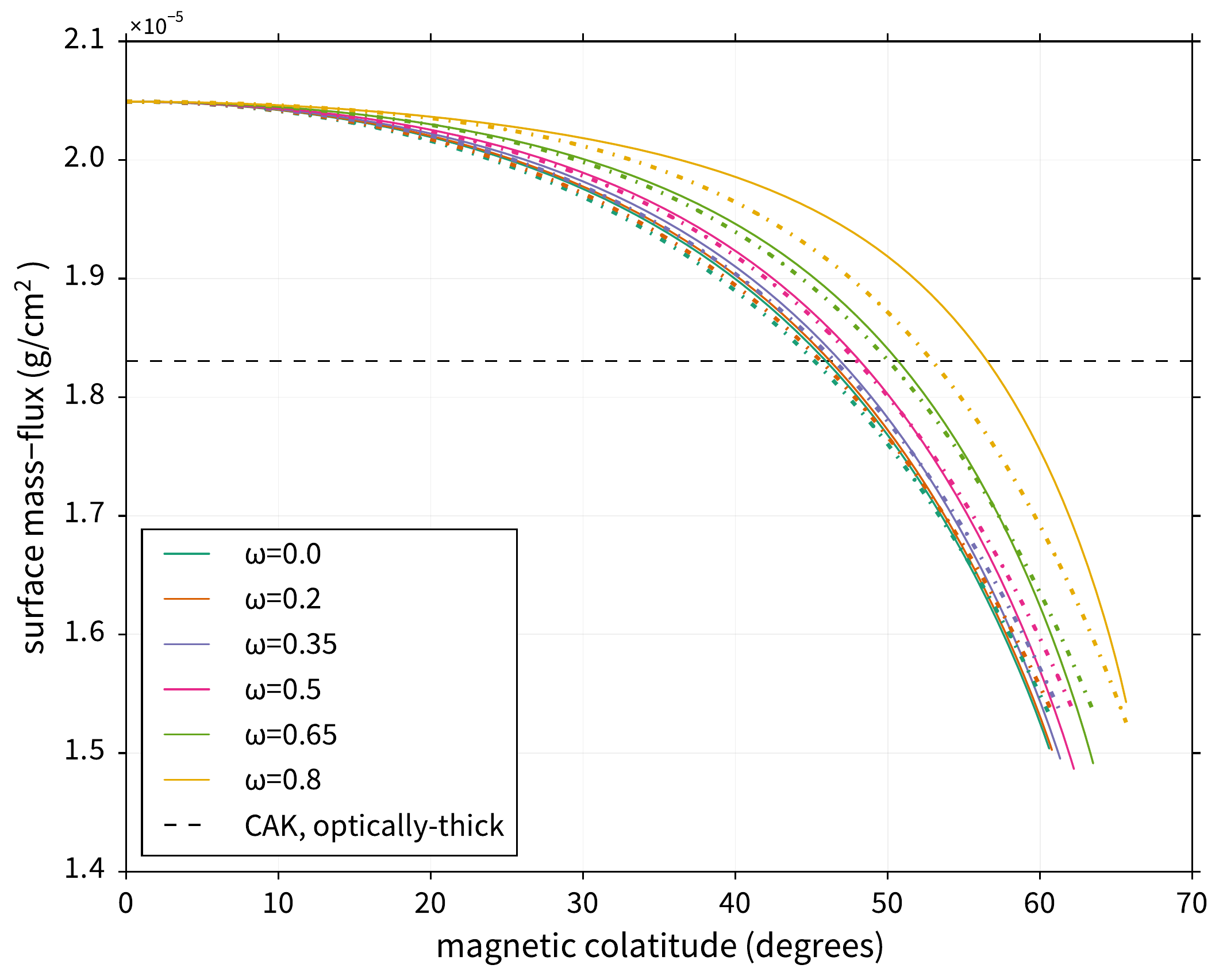}
\caption{\label{fig:M_flux_rot_unsat_O}Surface mass-flux ($\dot m_*$; solid line) as a function of line surface colatitude for an optically-thick line force and an O-type star. $\dot m_*$ is compared to the rotation scaling (dot-dashed line) derived in \autoref{eq:mflux_rot_scaling} and the CAK mass-flux (dashed line).}
\end{figure}

Since the critical radius is no longer constant with colatitude in the rotating cases, we will need to know the critical radius for each magnetic footprint (Section \ref{sec:rcrit}) in order to get precise estimates.
Interestingly, despite this dependence on \rcrit{}, we can still take OD04-type approximations to get a reasonable mass-flux estimate for different rotation rates independent of \rcrit{}!
We take $\psi_c(R_p/R_*)^3 \approx \mu_B$ ($\mu_B$ given in \autoref{eq:mub}) and $\theta_c\approx\theta_m$ such that $\aleph$ is evaluated at the stellar surface:
\begin{align}
\aleph_m \equiv 1 - \frac{12\cos^3\left[\frac{1}{3}(\pi+\cos^{-1}(\omega\sin\theta_m))\right]}{\omega\sin\theta_m}.
\end{align}
The resulting scaling relation is then
\begin{align}
\dot m_* &\approx\mdip{}(r_{c,\theta=0})\mu_B\aleph_m^{1-1/\alpha}.
\end{align}
As before, we correct for neglecting $\Phi$ by using the model-calculated \mtz{} instead of the approximation \mdip{}:
\begin{align}
\label{eq:mflux_rot_scaling}\dot m_* &\sim \mtz{}\mu_B\aleph_m^{1-1/\alpha}.
\end{align}
\autoref{fig:M_flux_rot_unsat} and \autoref{fig:M_flux_rot_unsat_O} show the fortunate agreement of \autoref{eq:mflux_rot_scaling} with our model calculations, despite the questionable approximations.
Again, we get larger differences between model and scaling for the faster rotation rates.

\subsection{With rotation in the general case}
\label{sec:mflux_rot_sat}
\begin{figure}
\includegraphics[width=0.49\textwidth]{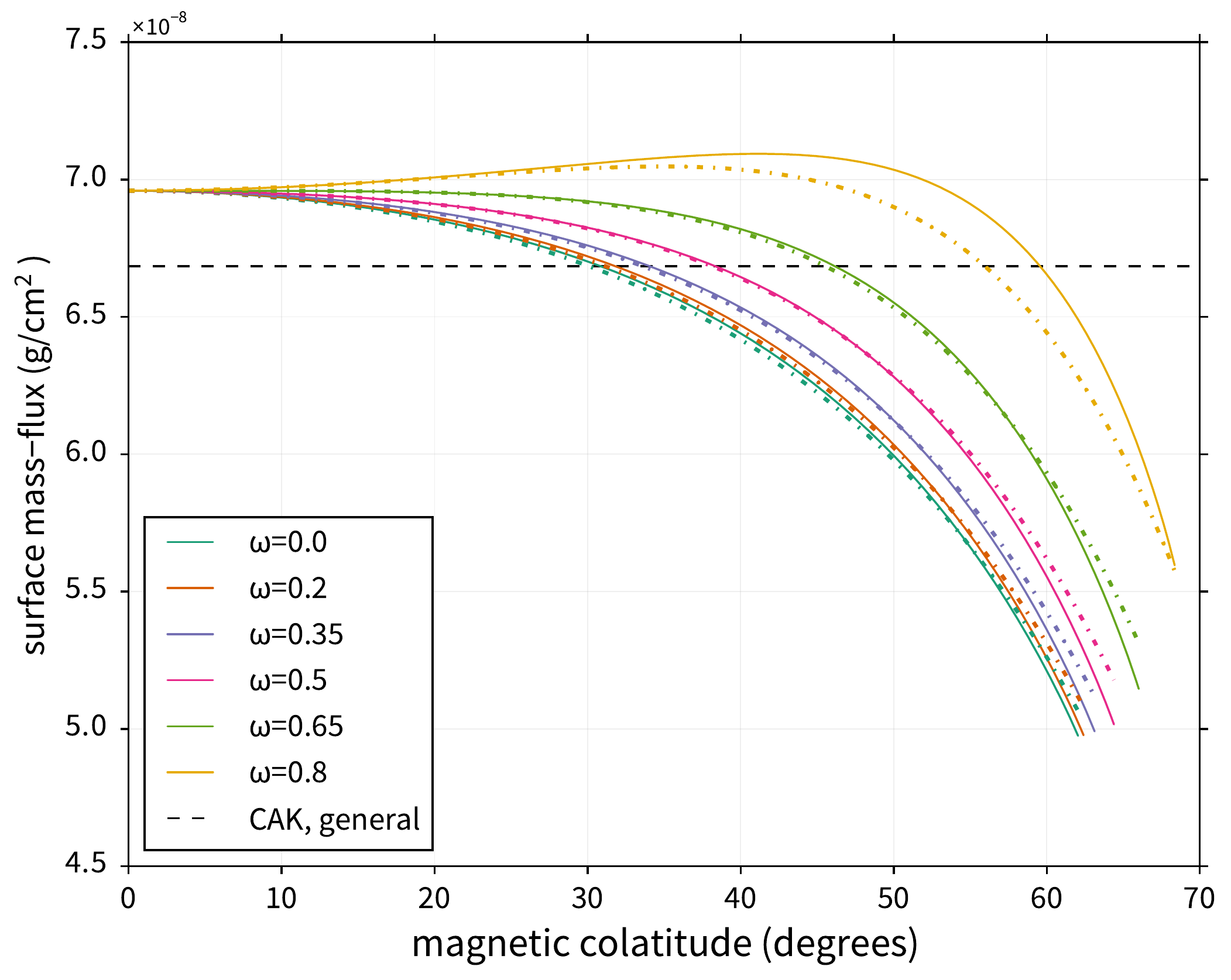}
\caption{\label{fig:M_flux_rot_sat}Surface mass-flux ($\dot m_*$; solid line) as a function of line surface colatitude for a general line force and a B-type star (\autoref{tab:star}). $\dot m_*$ is compared to the rotation scaling (dot-dashed line) derived in \autoref{eq:flux_scaling_rot_sat} and $\Sigma_0\mcak{}$, the corrected CAK mass-flux (dashed line).}
\end{figure}
Finally, we combine the effects of rotation and the OTC parameter.
With a similar derivation as the previous sections, we obtain
\begin{align}
\dot m_* \approx \mdip{}(r_c)\sigsatrot{}\left(\frac{R_p}{R_*}\right)^3 \aleph^{1-1/\alpha}\psi_c\sqrt{\frac{1+3\cos^2\theta_m}{1+3\cos^2\theta_c}},
\end{align}
where, again, \mdip{} is the base mass-flux from \autoref{eq:mdip}.
Unlike before, however, the OTC parameter now has a dependency on rotation:
\begin{align}
\label{eq:sat_rot_param}\sigsatrot{} = \frac{\left|1-\alpha-\left[\frac{1-\chi_0^{1/\alpha-1}\aleph^{1/\alpha-1}}{1-\chi_0^{1/\alpha}\aleph^{1/\alpha}}\right]\right|}{\alpha}
\end{align}
At zero rotation, $\aleph = 1$ and \sigsatrot{} reduces to \sigsatz{}.
With rotation, $\chi_0\aleph$ decreases and \sigsatrot{} moves towards 1.
Physically, this occurs because the rotation-boosted mass-flux in the wind further reduces the error from assuming an optically-thick \grad{}.
In more massive stars, rotation has less of an effect on the OTC parameter.
This is because the error from assuming an optically-thick wind is already small, so increasing the density in the wind does not have a relatively large effect.

We can simplify the scaling relation using OD04-type approximations:
\begin{align}
\dot m_*&\approx \mu_B\mdip{}(r_{c,\theta=0})\sigsatrotm{}\aleph_m^{1-1/\alpha},
\end{align}
where $\Sigma_\mathrm{rot,m}$ is calculated at the stellar surface (i.e. \autoref{eq:sat_rot_param} with $\aleph_m$ instead of $\aleph$).
We correct this for approximations made in Appendix \ref{app:mflux_scaling_sat} by using the model-calculated \mtzsat{} instead of $\sigsatz{}\mdip{}$:
\begin{align}
\label{eq:flux_scaling_rot_sat}\dot m_*\sim \frac{\Sigma_\mathrm{rot,m}}{\sigsatz{}}\mu_B\aleph_m^{1-1/\alpha}\mtzsat{}.
\end{align}

\begin{figure}
\includegraphics[width=0.49\textwidth]{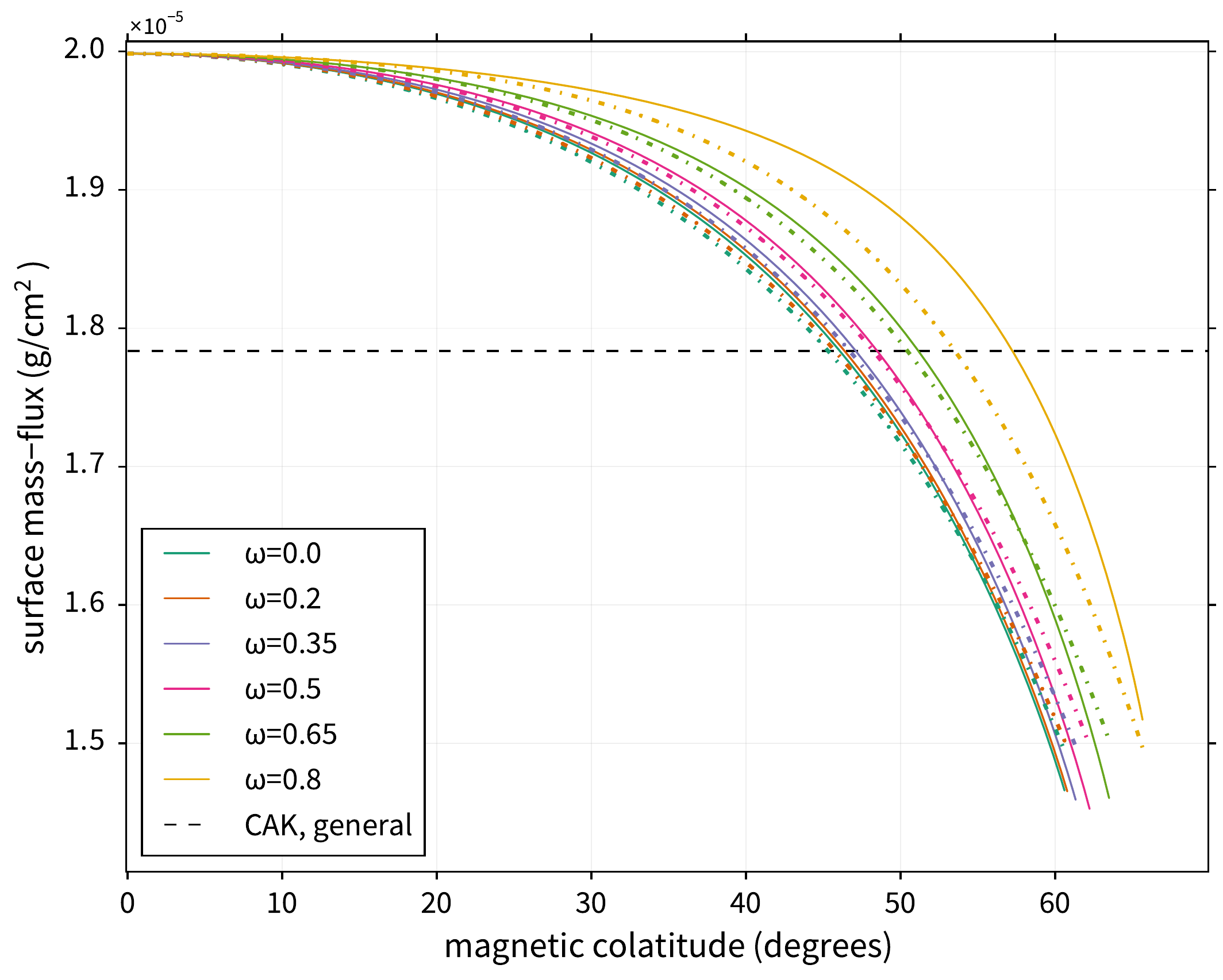}
\caption{\label{fig:M_flux_rot_sat_O}Surface mass-flux ($\dot m_*$; solid line) as a function of line surface colatitude for a general line force and an O-type star. $\dot m_*$ is compared to the rotation scaling (dot-dashed line) derived in \autoref{eq:flux_scaling_rot_sat} and $\Sigma_0\mcak{}$, the optically-thin corrected CAK mass-flux (dashed line).}
\end{figure}

\autoref{fig:M_flux_rot_sat} and \autoref{fig:M_flux_rot_sat_O} show the good agreement of our scaling (\autoref{eq:flux_scaling_rot_sat}) with our model calculations for both the B- and O- type stars.
Again, we get larger differences between the model and the scaling relation at faster rotation rates.
At $\omega = 0.8$, we get about a 3\% difference for the B star and 4\% for the O star.

\section{Critical point locations}
\label{sec:rcrit}
The critical point location, \rcrit{}, depends not only on the boundary condition, but also on the inherent properties of the magnetosphere and wind.
In CAK theory, this location sets the value of the surface mass-flux, so anything that moves this point influences the amount of material being accelerated off the stellar surface.
There has been some discussion in the literature about the physicality of the CAK critical point (e.g. \citealp{lamers99, lucy07}) and its validity in setting the critical mass-flux. 
These authors prefer using the sonic point to set the critical mass-flux (e.g. in the models of \citealp{vink00}).
For now, we defer discussion of this issue to future studies.

\begin{figure}
\includegraphics[width=0.49\textwidth]{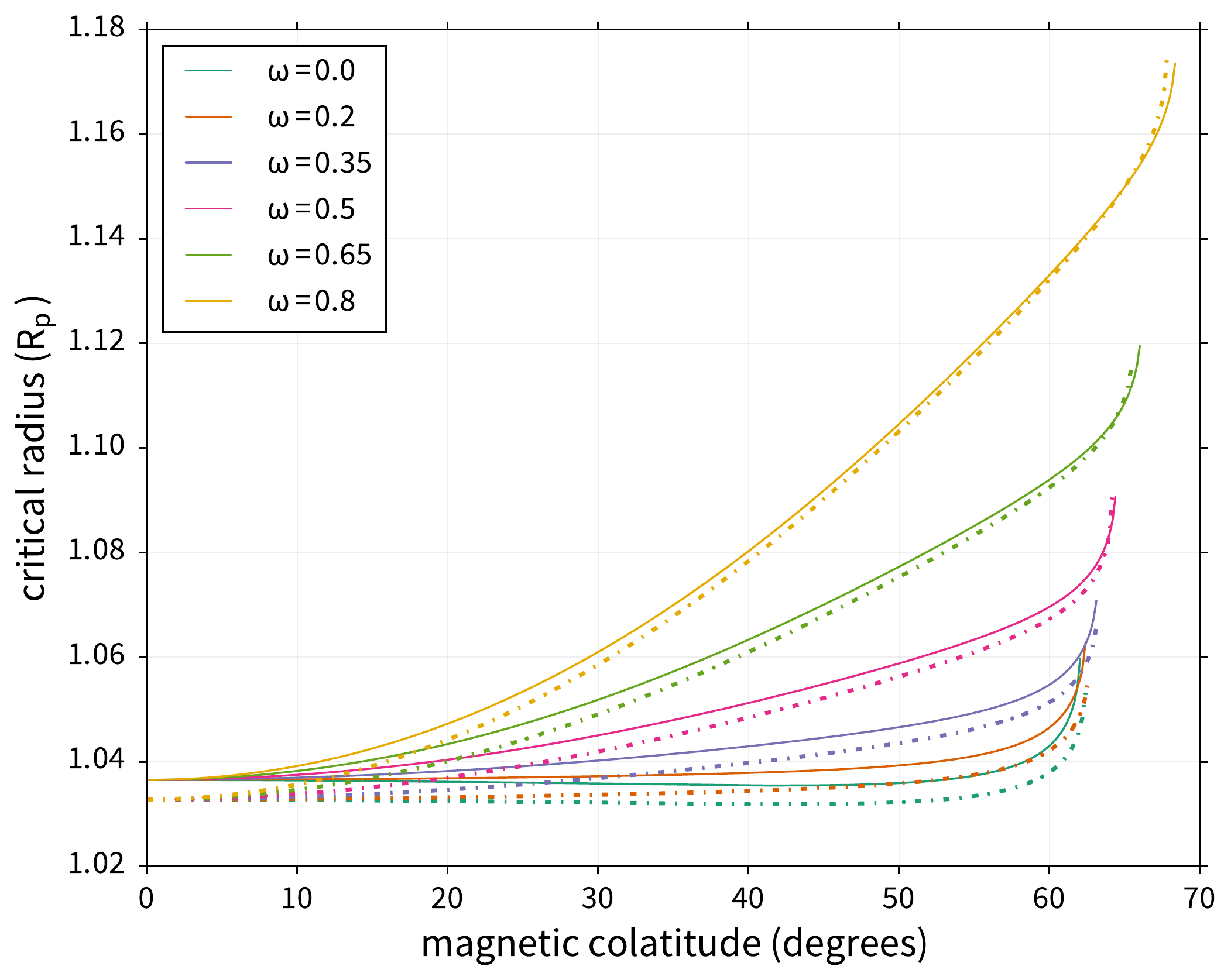}
\caption{\label{fig:crit_r}Critical point location \rcrit{} in the general (solid) and optically-thick (dot-dashed) cases as a function of critical rotation fraction $\omega$ and dipole line surface colatitude $\theta_m$ for a B-type star. In general, both rotation and the optically-thin correction move the critical point away from the center of the star, though the rotation effect is caused by the increased stellar radius of an oblate star (\autoref{fig:crit_r_surf}). Similar results for the O star not shown.}
\end{figure}

In the point-star zero sound-speed limit, CAK found that the entire wind is critical; this degeneracy means that the critical radius ($r_c$) is ill-defined for this case.
Including the small sound-speed term barely breaks this degeneracy, but the finite-disk correction allows $r_c$ to be well-defined by allowing the critical velocity, its derivative, and the mass-loss rate to vary with radius \citep{kudritzki89, madura07}.
This results in only one radius which satisfies the critical and boundary conditions simultaneously.

Interestingly, dipole divergence also breaks this degeneracy!
Although it does not change the critical velocity and its derivative (\autoref{eq:y_c_approx}), the faster-than-spherical expansion induces a mass-flux dependency on the critical radius (\autoref{eq:mdip}).
There is then only one radius which allows a self-consistent critical mass-flux.
This critical location is close to the star ($r_c \approx 1.033 R_p $ for the B-star; $\approx 1.054 R_p$ for the O-star), much like the finite-disk corrected critical radius ($\approx 1.02 R_p$ in \citet{pauldrach86} with different stellar properties for their O-star).

Next, we study two factors which influence the critical location, the optically-thin correction and stellar rotation.
Both these effects work in the same manner, causing the smaller root of \autoref{eq:eom} and, equivalently, the wind acceleration in the subcritical region to decrease.
Even though these factors also lower the critical velocity (\autoref{eq:gen_unsat_u}), they have different effects on \rcrit{}.
The optically-thin wind takes a longer distance to accelerate from the boundary to the critical point, which pushes \rcrit{} out.
However, rotation lowers the critical velocity sufficiently enough that the wind is able to accelerate over a slightly shorter distance, pulling \rcrit{} in relative to the stellar surface.
Stellar oblation, though, will push the critical radius out relative to the center of the star.

\begin{figure}
\includegraphics[width=0.49\textwidth]{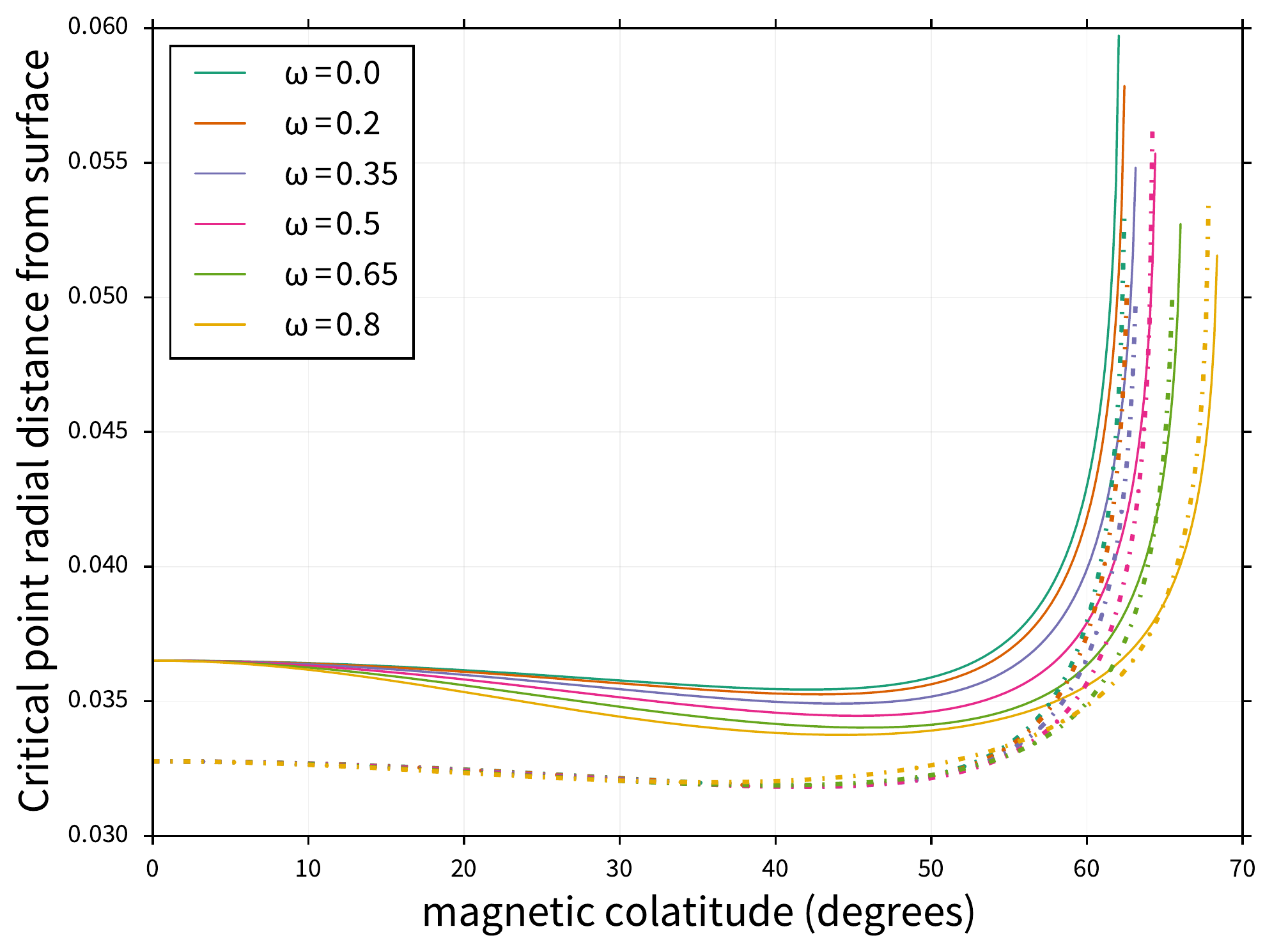}
\caption{\label{fig:crit_r_surf}Critical point location realative to the stellar surface (\rcrit{} - $R_*$) in the general (solid) and optically-thick (dot-dashed) cases as a function of critical rotation fraction $\omega$ and dipole line surface colatitude $\theta_m$ for a B-type star. The optically-thin correction tends to move the critical radius away from the star, but rotation moves the critical radius slightly closer to the star. Similar results for the O star not shown.}
\end{figure}

Our results for a dipole magnetosphere (\autoref{fig:crit_r}/\autoref{fig:crit_r_surf}) show how \rcrit{} moves out due to rotation and the OTC parameter.
We note that the increase in critical radius due to rotation is almost entirely caused by the stellar oblation.
In fact, we see that the radial distance of the critical point from the stellar surface is nearly constant with colatitude and rotation.
For more massive stars, the error from the optically-thick assumption is reduced (see Section \ref{sec:mflux_sat_zero} for discussion), so there is a smaller difference in \rcrit{} between the general and optically-thick cases than for later-type stars.

Finally, we note that the critical radius does not exist for every field line.
In \autoref{fig:crit_r}, we see a clear tendency for the footprint colatitude to have a limit.
Another interesting aspect is that different rotation rates have different $\theta_m$ limits, though as we will see, this limit may actually be determined by the field line shell parameter ($L$).
The starting footprint locations of our six model magnetospheres (one for each rotation fraction) are identical, but the different stellar surfaces resulting from rotation (c.f. \autoref{eq:nhat}) yield different $L$ values for a single $\theta_m$. 
Faster rotation results in a greater $L$ for a given surface colatitude.
When we evaluate $L_\mathrm{min}$ corresponding to the maximum $\theta_m$ for each rotation rate, we get similar values: $L_\mathrm{min}\approx 1.27-1.3 R_*$ for a B-star, $L_\mathrm{min}\approx 1.32-1.33 R_*$ for a O-star, with slight differences between the general and optically-thick cases.
We stress that these limits are only approximate due to the division of our model stellar surface into 500 discrete magnetic footprints.

It is uncertain exactly why this limit, if there is one, exists.
For lines close to this limit, there do exist possible critical points, but none satisify the boundary condition.
This occurs because the magnetic tension due to line tilt (represented by $\psi$) neuters the wind acceleration so that the flow cannot pass through the critical point for the given boundary condition.
For lines much closer to the equator, there are no possible critical points for any boundary condition since $\Phi^2<0$ (Section \ref{sec:topo}).

\section{Velocity structure}
\label{sec:term_veloc}
Here, we focus on the wind velocity as it is accelerated along a dipole field line and qualify its behavior.
\subsection{Is there a beta-velocity law?}
In the CAK zero sound-speed limit, the radial velocity structure is found to be $v = v_\infty(1-R_*/r)^\beta$, with $\beta = 1/2$ for this specific case.
With the finite-disk correction, the velocity can still be well-represented by this beta-velocity law \citep{pauldrach86}, which greatly simplifies analytic considerations of the finite-disk correction factor by allowing the factor to be represented as an explicit spatial function (e.g. \citealp{madura07}).
However, for a magnetosphere, there are two problems with assuming a beta-velocity law: the terminal velocity and the effect of rotation.

The faster-than-spherical divergence of the field will lead to higher terminal velocities \citep{owocki04} than for the spherically-diverging case.
With the escape velocity defined as $v^2_\mathrm{esc} = 2(1-\gammael{})GM_*/R_p$, we calculate the polar $v_\infty\approx 1.97~v_\mathrm{esc}$ ($5.9~v_\mathrm{esc}$) for our B-type (O-type) OTC wind.
We note that observed terminal velocities of magnetic stars are unlikely to ever reach these limits since the faster-than-spherical divergence of the wind will not continue indefinitely through space.
For comparison, the modified CAK terminal velocity for our O-star is 3000 km s$^{-1}$ $=4.23~v_\mathrm{esc}$ \citep{uddoula14}; however, this is based on the optically-thick \grad{} which does not properly reduce to the optically-thin limit at low densities.
Observations of non-magnetic stars give lower terminal values $v_\infty\approx 1.4 v_\mathrm{esc}$ for B-stars and $v_\infty\approx 2-3~v_\mathrm{esc}$ for O-stars \citep{kudritzki00}.

Of course, closed field lines cannot have terminal velocities as they do not extend to infinity.
The ``terminal velocities'' we find in this paper for each field line are merely best-fit parameters used to characterize the velocity behavior.
To quantify the behavior, we use nonlinear least squares to fit individual beta-velocity laws to the numerically-calculated velocity structure of each line.
The velocity structures are calculated from the critical point to the apex, ignoring any possible shocks.
Thus, our calculated beta-velocity laws will accurately describe the wind velocity up to the shock, though not past it.

\begin{figure}
\includegraphics[width=0.49\textwidth]{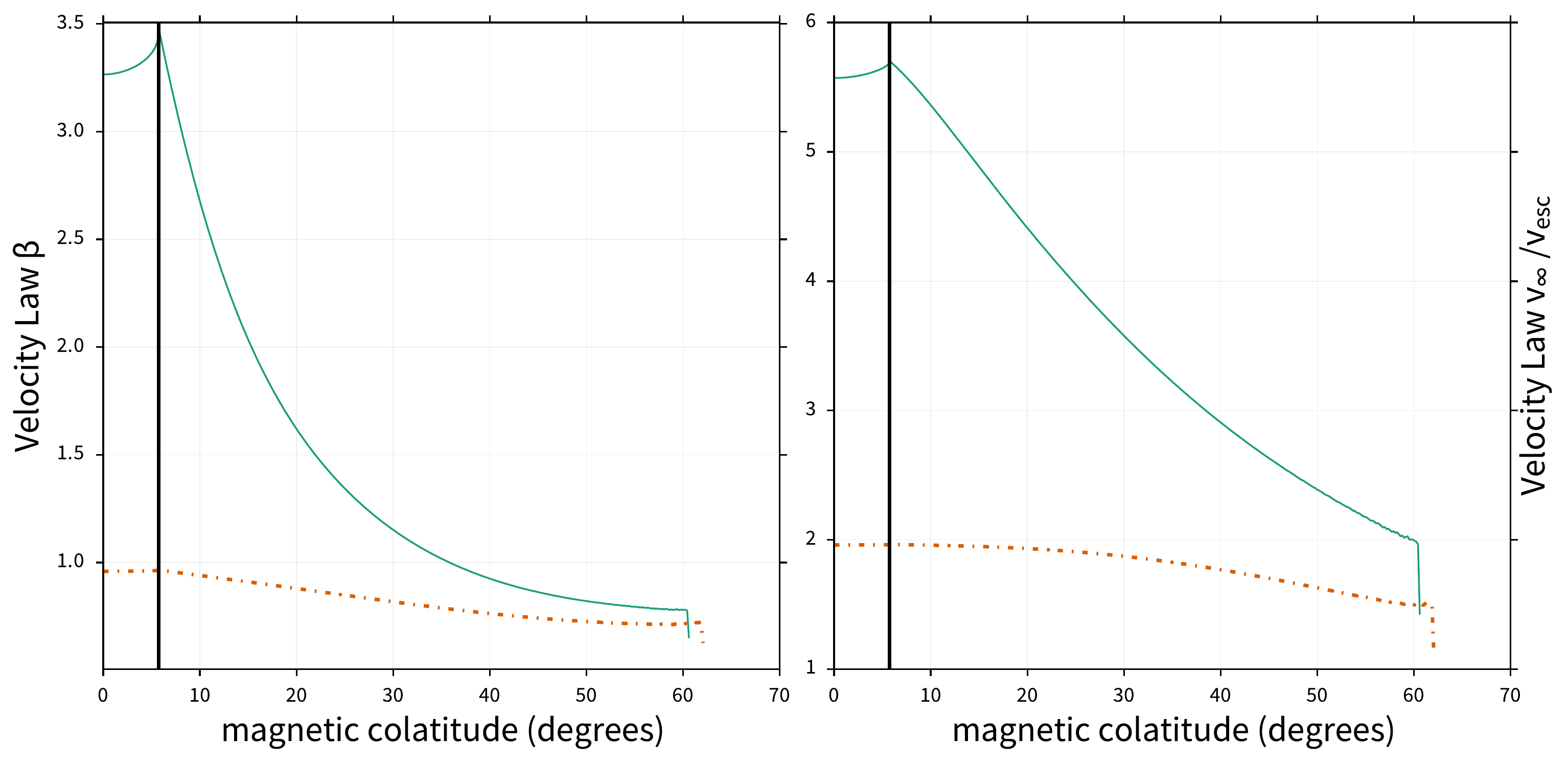}
\caption{\label{fig:Veloc_fit}Best-fit velocity-law betas (left) and corresponding terminal velocities (right) for our non-rotating O-star (solid) and B-star (dot-dashed) models. Fit parameters were determined using nonlinear least-squares fitting. The solid black line represents the surface colatitude for $L = 100~R_p$, our truncation radius; results from more polar colatitudes should be ignored since those model lines were truncated well before reaching an apex.}
\end{figure}

We find that that there is no easily-defined global beta-velocity law for the magnetosphere. 
Instead, each line has an independent velocity structure which depends on its own geometry.
For a non-rotating magnetic dipole, the best-fit value of both $v_\infty$ and $\beta$ varies throughout the magnetosphere (\autoref{fig:Veloc_fit}).
As one moves toward more polar colatitudes, this best-fit $v_\infty$ approaches the asymptotic limit of the straight-line dipole terminal velocity.
The magnetic field geometry also affects how quickly the wind accelerates; higher tilt relative to the surface reduces both the ``terminal velocity'' and how long it takes the wind to reach that limit (as represented by a decreasing $\beta$).

Interestingly, the best-fit $\beta$ values are quite large for our model O-star ($\beta > 2$) and not within the usual non-magnetic range $0.7 \lesssim \beta\lesssim 1$ \citep{kudritzki89}.
The model B-star shows similar behavior, though the $\beta$ range fits better with non-magnetic values.
Both cases imply that the field-line tilt has a large effect on throttling the wind acceleration even as the dipole divergence works to boost it.

We note that that finite-disk effect will change these velocity results. 
Compared to the point-star CAK model, the finite-disk effect leads to a higher terminal velocity because less mass is driven off the star and this lower-density wind sees more of the stellar surface as it accelerates out.
However, these modified CAK models use the optically-thick version of \grad{}, which artifically boosts this low-density acceleration and leads to higher terminal velocities.
We will implement the finite-disk correction in Paper II and characterize its effect on the velocity structure there.

\begin{figure}
\includegraphics[width=0.49\textwidth]{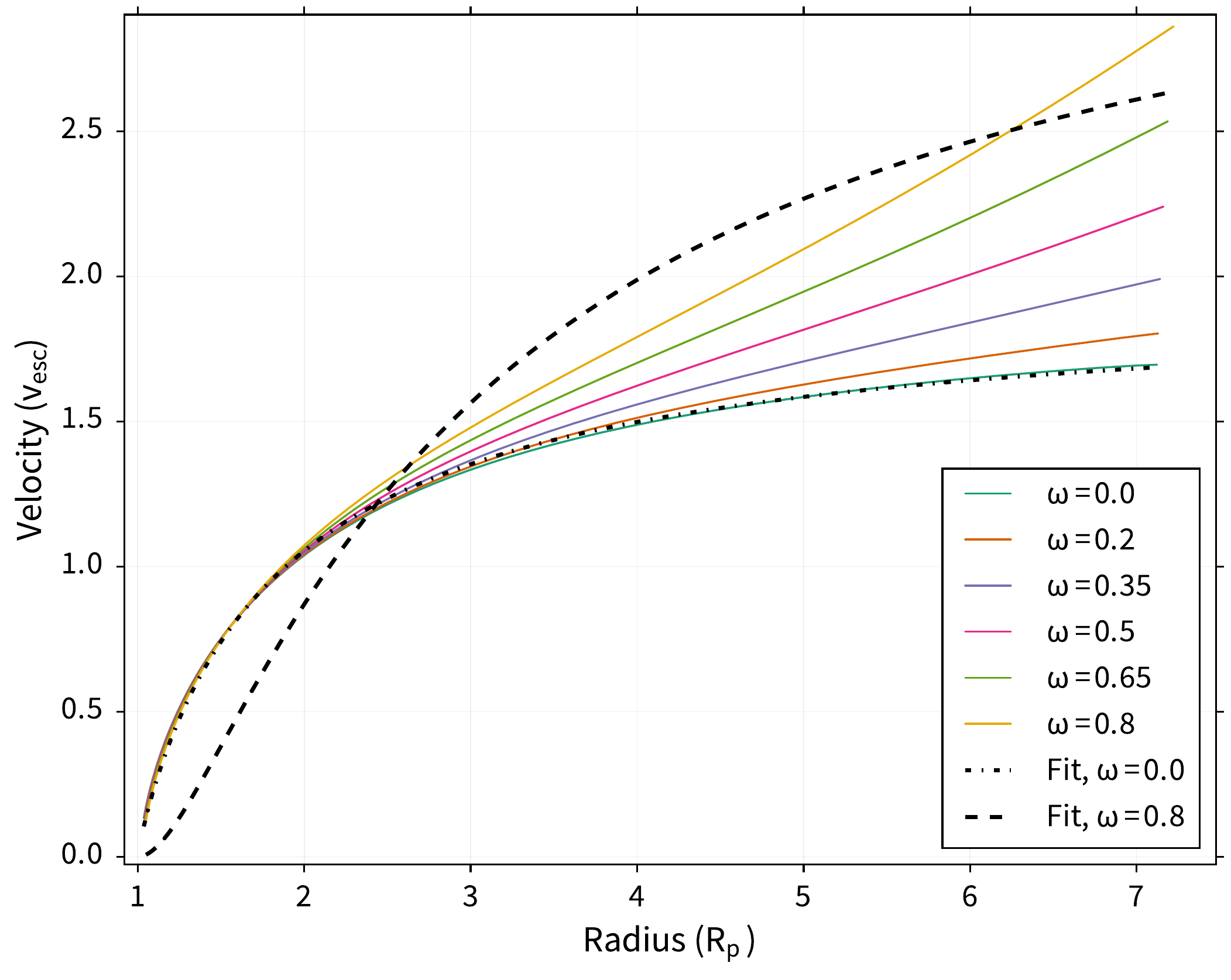}
\caption{\label{fig:veloc_rot} Velocity in terms of the stellar escape velocity along one model B-star magnetic dipole line ($\theta_m = 22^\circ$; $L= 7.12-7.22~R_p$) for several stellar rotation fractions $\omega$. The best-fit beta-laws from nonlinear least-squares fitting for zero rotation (dot-dashed) and high rotation (dashed) are also shown.}
\end{figure}

The second issue with a global magnetospheric beta-velocity law is the acceleration from rigid-body rotation. 
Beyond a certain point, centrifugal acceleration will exceed the inward gravitational pull and cause the wind to accelerate; there is no asymptotic limit.
This breaks the beta-velocity law assumption and makes it difficult to characterize the velocity structure with a general equation (\autoref{fig:veloc_rot}).

We note that for the more equatorial lines in our model, the wind did not actually accelerate all the way from the critical point to the line apex.
Instead, a kink solution occurs since the equation of motion (\autoref{eq:eom}) stops admitting positive roots for the velocity derivative at some location while still allowing the negative roots (c.f. \citealp{cranmer96a}; \citealp{madura07}).
In order to fit a beta-velocity law to these lines, we ignored the deceleration after the kink and only fit the portion of the line from the critical point to the deceleration point.
This implicity assumes that shocks along each line will prevent the wind from ever reaching a kink, so these fits will represent the behavior of the wind velocity up to the shock.

\subsection{Closure radius}
\label{sec:closure}
There is always a struggle between the wind and the magnetic field within magnetospheres.
As quantified by $\eta_*$ (\autoref{eq:eta_star}), the field dominates the wind if its energy is larger than the wind kinetic energy.
On the other hand, the wind will escape if its velocity exceeds the local Alfv\'en velocity $v_A = B/\sqrt{4\pi\rho}$.
We can thus understand the Alfv\'en radius ($R_A$), the typical length scale of the magnetosphere, as the point where $v=v_A$.
Additionally, we can approximate the maximum extent of closed loops in the magnetosphere, the closure radius ($R_c$), as the $L_\mathrm{shell}$ for which $v_\mathrm{apex} = v_A$.
This will only be a lower bound in the context of this model ($\eta_*\to\infty$) since the shocks produced by colliding wind flows will not allow the wind to fully accelerate all the way to the line apex.

We can compare this to the MHD-derived closure radius scaling \citep{uddoula02}
\begin{align}
\label{eq:r_close}R_c \approx R_* + 0.7(R_A - R_*),
\end{align}
with the dipole Alfv\'en radius given by
\begin{align}
\frac{R_A}{R_p} \approx 0.3 + \eta_*^{1/4}
\end{align}
characterizing the maximum radius at which the magnetic field still dominates the wind.

\begin{figure}
\includegraphics[width=0.49\textwidth]{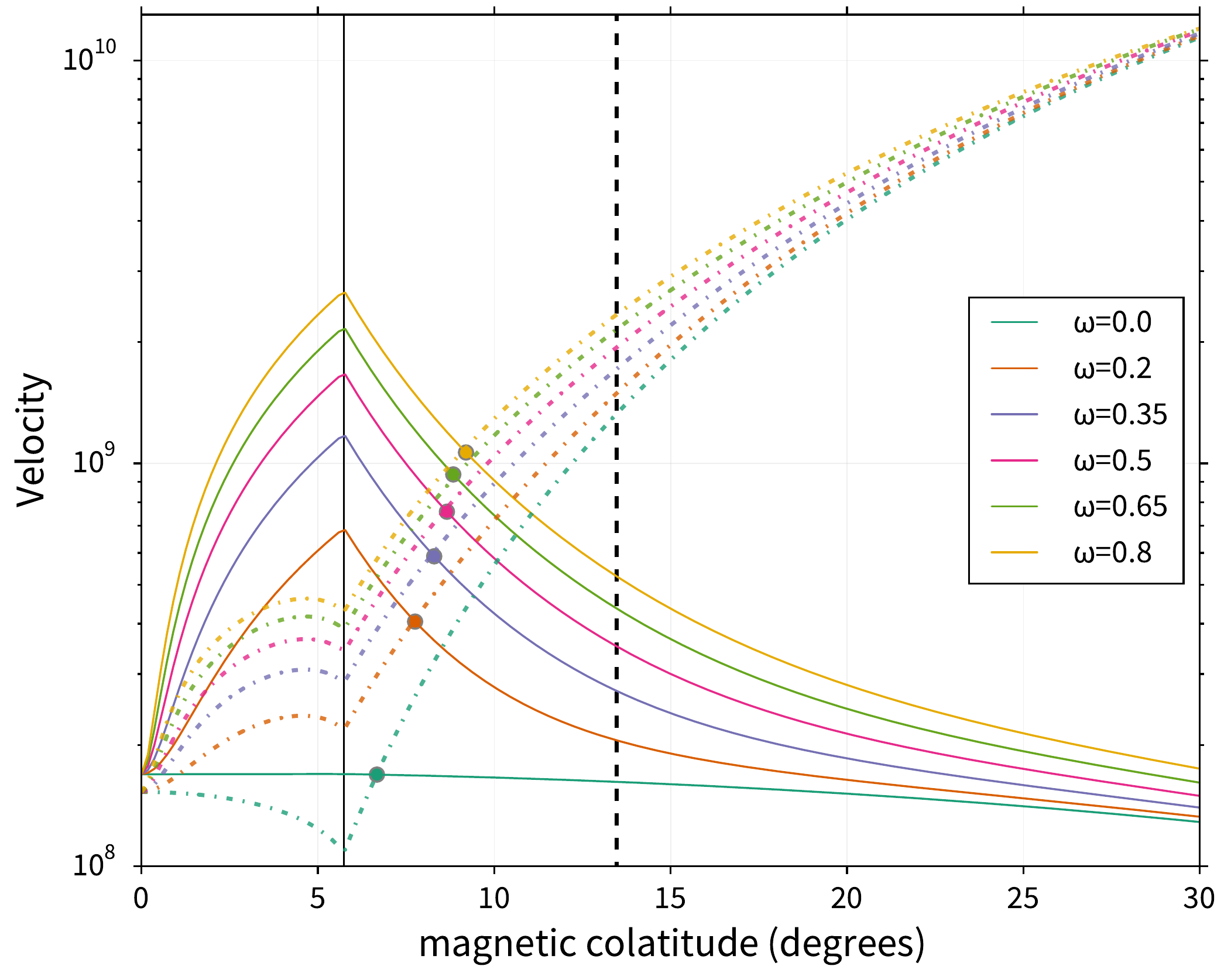}
\caption{\label{fig:close_B} Comparison of wind velocity (solid) vs. Alfv\'en velocity (dot-dashed) at apex of each magnetic field line for our optically-thin corrected B-star model. Colored dots indicate approximation of closure colatitude for each rotation rate. The dashed line indicates the non-rotating MHD-predicted closure colatitude. The solid black line represents the surface colatitude for $L = 100~R_p$, our truncation radius; results from more polar colatitudes should be ignored since those model lines were truncated well before reaching an apex.}
\end{figure}

For our model stars, we assume the same non-magnetic $v_\infty$ and $\dot M$ as the previous subsection.
We take the $\sigma$ Ori E-like value $B_\mathrm{eq} = 5500 \unit{G}$ for our B-star and take $B_\mathrm{eq} = 1857.5 \unit{G}$ such that $\eta_* = 100$ for our O-star.
This results in a MHD-predicted closure radius $R_c = 18.4 R_p$ ($2.72 R_p$) for the B-star (O-star).

Not surprisingly, we obtain larger closure radii than MHD predictions (\autoref{tab:r_close}).
This is due to our rigid-field assumption; in reality, the wind will stretch out the polar field lines radially \citep{uddoula02} and accelerate more rapidly when $\psi$ moves towards unity.
The resulting faster velocities would move $R_c$ towards the MHD-approximated closure radius.
This effect is more important for stars with smaller $\eta_*$, since the weaker confinement will allow the wind to have more effect on the magnetic topology.
However, we note that the MHD simulations of \citet{uddoula02} only considered $\eta_* \lesssim 100$; our model B star has $\eta_*\approx 4\e{5}$.
Since we are not able to efficiently simulate these B star magnetospheres with MHD codes, it is unclear at the moment how important radial stretching will be for such strong magnetic fields.

\begin{table}
\centering
\caption{\label{tab:r_close} Estimated closure radii (in units $R_p$) for the model B-star centrifugal magnetosphere and O-star dynamical magnetosphere at different rotation rates. The MHD scaling estimate (\autoref{eq:r_close}) is also included for comparison.}
\begin{tabular}{cccccccc}
\hline\hline
Type & $\omega = 0.0$ & 0.2 & 0.35 & 0.5 & 0.65 & 0.8 & MHD\\\hline
$B$ & 74.1 & 54.9 & 48.0 &44.2 & 42.4 & 39.2 & 18.4\\
\hline
$O$ &5.02 &4.96 &4.91 &4.87 &4.78 & 4.65 & 2.72
\end{tabular}
\end{table}

We also see a clear trend of rotation boosting both the apex wind and Alfv\'en velocities.
The Alfv\'en velocities increase since, by the conservation of mass, the faster wind velocities result in lower densities.
The overall effect is to produce smaller closure radii at faster rotation rates.

\begin{figure}
\includegraphics[width=0.49\textwidth]{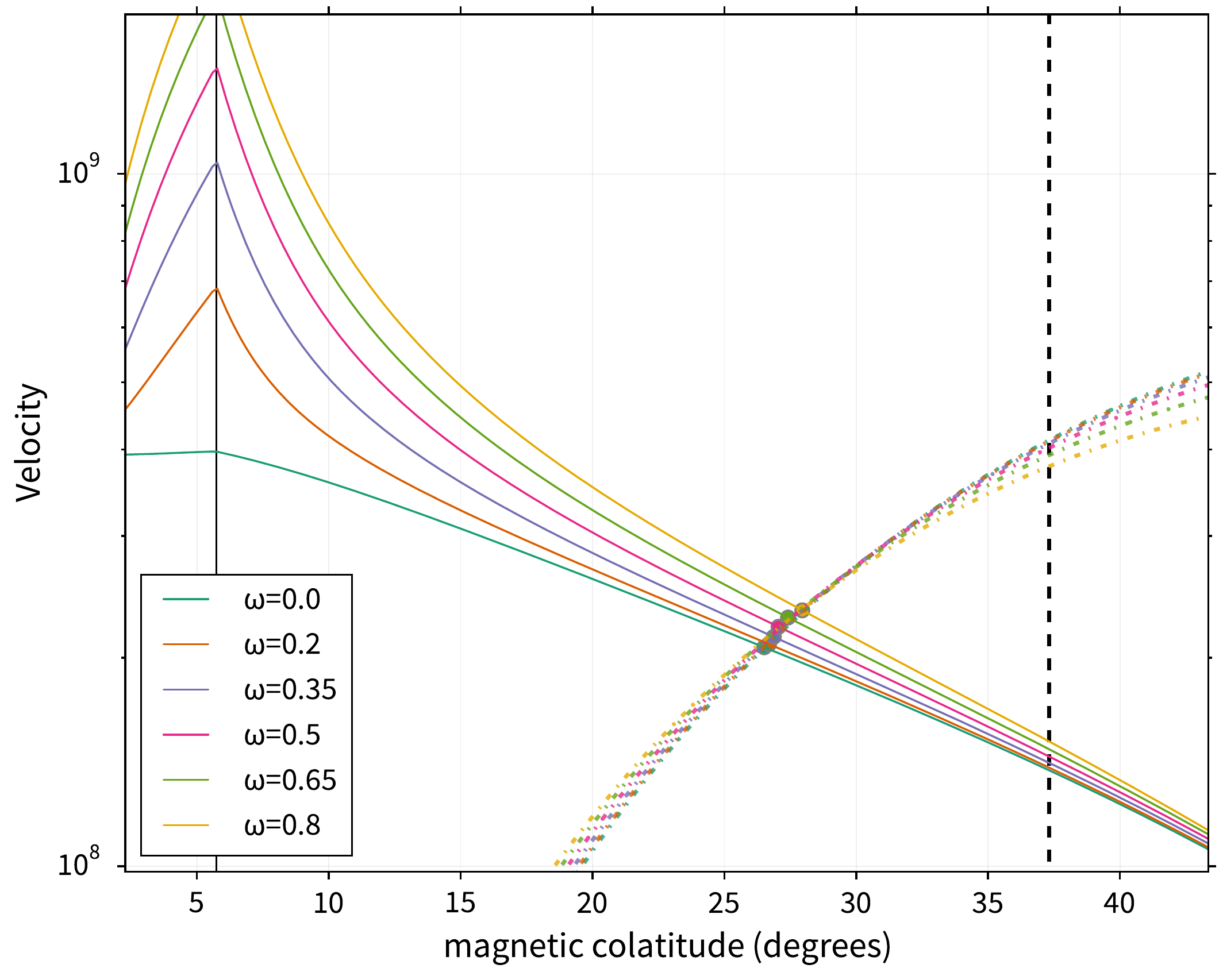}
\caption{\label{fig:close_O} Same as \autoref{fig:close_B}, except for our O-star model.}
\end{figure}

\section{Global mass loss}
\label{sec:mass_loss}

\begin{table*}
\centering
\caption{\label{tab:B_massloss}Mass-loss rates (in units of $ 10^{-9}~M_\odot/\unit{yr}$) for a B-type star ($\eta_* = 4.28\e{5}$) as calculated from our model, assuming infinite magnetic confinement where applicable. ``No $B$'' indicates a CAK-type mass-loss rate calculated from a non-rotating radial flow with spherical divergence. The other mass-loss rates are calculated from a dipole magnetosphere with the given rotation fraction $\omega$. ``Optically-Thick'' indicates the mass-loss calculated from using the optically-thick \grad{}; the rest use the ``General'' \grad{}. ``True'' is the mass-loss into open field lines ($L>R_c$), $R_c$ given by the MHD-estimated value (\autoref{eq:r_close}). ``Disk'' is the mass-loss into field lines with a centrifugally supported disk ($R_K < L < R_c$). ``Effective'' is the mass-loss which does not fall back to the star; it is the sum of the ``True'' and ``Disk'' mass-loss rates. Numbers in parentheses next to a mass-loss rate represent the ratio of that particular rate to the ``General'' mass-loss at its rotation fraction $\omega$.}
\begin{tabular}{lccccccc}
\hline\hline
 &No $B$ & $\omega = 0.0$ & 0.2 & 0.35 & 0.5 & 0.65 & 0.8 \\
\hline
Optically-Thick & 1.81 & 1.00 & 1.01 & 1.02 & 1.04 & 1.08 & 1.12\\
General &1.29 & 0.718 & 0.724 & 0.738 & 0.762 & 0.797 & 0.845\\
True &... &0.036(0.05) &0.036(0.05) &0.036(0.049) &0.037(0.049) &0.037 (0.046) &0.037(0.044)\\
Disk &... &... &0.21(0.29) & 0.33 (0.45)& 0.46 (0.6) & 0.61 (0.77)&0.8 (0.95)\\
Effective &... & 0.036 (0.05) & 0.24(0.33) &0.37(0.5) &0.49 (0.64)&0.64 (0.8)& 0.84 (0.99)\\
\hline
\end{tabular}
\end{table*}

Here, we study the effect of the dipole field on the global mass-loss rate for different rotation rates.
We can find the global mass-loss rate, $\Mglo{}$, by integrating the mass-flux over the stellar surface:
\begin{align}
\label{eq:global_Mdot}\Mglo{} = \int \dot m_r~dA = 2\pi\int R_*^2\mu_B\dot m_* d\mu_*~,
\end{align}
with $\mu_*\equiv\cos\theta_m$.
Since not every field line has a calculated critical mass-loss rate, we will assume that the various scaling relations for $\dot m_*$ derived in Section \ref{sec:mass_flux} hold for the entire stellar surface.
For the non-magnetic case, we take the mass-flux for a straight line with spherical divergence ($\dot m_{B=0}$) and integrate over the stellar surface:
\begin{align}
\dot M_\mathrm{global, B=0} = \dot m_{B=0} \int dA~ = 4\pi R_p^2\dot m_{B=0}.
\end{align}

We stress that the mass-loss rate in \autoref{eq:global_Mdot} is not a ``true'' mass-loss rate; the plasma flowing along closed field lines does not easily escape the magnetosphere \citep{uddoula08, townsend13}.
In our model, our assumption of infinite magnetic confinement means that, technically, none of the stellar wind escapes the magnetic field.
Despite this, however, \Mglo{} is still an useful value to calculate, as it will give a better estimate for the wind magnetic confinement parameter \citep{uddoula02}, depending on rotation.
The amount of mass that escapes into the interstellar medium depends on the closure radius (\autoref{eq:r_close}), which itself depends on the stellar magnetic field strength.
Lines with a shell parameter $L > R_c$ are considered ``open'' for the purposes of calculating a ``true'' mass-loss rate.

We also calculate a ``disk'' mass-loss rate into a centrifugally-supported disk.
Such disks are created because maintaining rigid-body rotation away from the star eventually leads to a balance between the gravitational and centrifugal forces at the Kepler radius (e.g. \citealp{uddoula08})
\begin{align}
\label{eq:R_kep}R_K = \frac{GM_*}{v_\phi^2} = \omega^{-2/3}R_p.
\end{align}
For lines inside the Kepler radius, the lack of centrifugal support results in a pattern of outflow and infall that leads to a long-term average mass-loss of zero \citep{uddoula02}.
However, for lines outside the Kepler radius, the wind never falls back to the star and instead remains in a disk, suspended away from the star.
Combining the ``disk'' and ``true'' rates can give us an ``effective'' mass-loss; the star loses mass if it will not return to the surface, either because it settles in a magnetospheric disk or escapes through open field lines.

The results for our model magnetospheres are presented in \autoref{tab:B_massloss} (B-star) and \autoref{tab:O_massloss} (O-star).
For zero rotation, the ratio between the general and optically-thick \Mglo{} is well explained by the OTC parameter, \sigsatz{}.
For our B-star, we calculate a ratio $7.18\e{-10}/1\e{-9}\approx 0.718$, which compares well to our approximated $\sigsatz{}\approx 0.725$.
For the O-star, the model \Mglo{} ratio is $3.67\e{-6}/3.77\e{-6}\approx 0.973$, which fits with our \sigsatz{}$\approx 0.974$.
Additionally, the ratio between the CAK-type mass-loss rates (``No B'' in \autoref{tab:B_massloss}/\autoref{tab:O_massloss}) between the general and optically-thick cases can be approximated by \sigsatz{}.
The reason why the ratios differ from the actual value of \sigsatz{} is because \rcrit{} is different between the general and optically-thick cases, leading to different base mass-fluxes (\autoref{eq:mdip}).
Nonetheless, multiplying the base mass-loss by the OTC parameter gives an excellent approximation.

The effective mass-loss rates for both our B and O stars imply that most of the plasma falls back to the star at low rotation (about 2/3rds at $\omega = 0.2$) and nearly none of it falls back at high rotation (1\% at $\omega = 0.8$).
This leads to mass-loss of about 20-65\% of the non-magnetic, non-rotating CAK value.

Since we use \citet{gayley95}'s $\bar Q$ parameterization for the line-acceleration, we must be careful when comparing calculated mass-loss rates with other models which use the more traditional CAK force multipler paradigm \citep{abbott82}.
\citet{puls00} show that using $\bar Q$ in \grad{} instead of the CAK $k$ parameter requires an ansatz that does not hold for $T_* < 35000 K$.
In our model, assuming a wind temperature equal to the stellar effective temperature means that B-star winds will be below this cutoff; the result is an overestimated mass-loss rate (c.f. Table 2 in \citealp{puls00}) by about a factor of 2.
Future studies will need to determine the wind temperature to check the validity of the $T=T_\mathrm{eff}$ assumption.

Further improvements to our global mass-loss rates will require consideration of the finite-disk effect.
We will implement this term in Paper II, but for now we can estimate the finite-disk corrected \Mglo{} by dividing our results by two.

\section{Summary and future work}
In this paper, we presented a critical point analysis of the Arbitrary Rigid-Field Hydrodynamic Equations, which represent a CAK-type wind within an arbitrary, infinitely-strong magnetic field.
This differs from the usual CAK wind model by including the proper optically-thin maximum line-force,  a rigid-body centrifugal acceleration, and a dipole areal divergence.
After finding the general critical point values for the mass-flux, velocity and velocity derivative, we confirmed that they reduced to the proper values for a traditional CAK wind, i.e. a non-rotating, non-optically-thin corrected, radial flow with spherical divergence.
These benchmarked general critical point equations were then applied to an aligned magnetic dipole field in order to calculate critical point locations and surface mass-fluxes.
By integrating from these critical point locations, the velocity structure within the magnetosphere was quantified and studied.
Finally, we obtained global mass-loss rates and found that the dipole field effectively reduces the overall mass-loss to 20-65\% of the non-magnetic, non-rotating CAK value.

The key results are summarized as follows:
\begin{enumerate}
\item We are able to approximately confirm the \citet{owocki04} scaling for the influence of a magnetic dipole on the surface mass-flux, $\dot m_r\approx \mu_B^2\mcak{}$. 
While this scaling does not need much improvement, we provide a more accurate scaling equation (\autoref{eq:mflux_zero_unsat}) and detail which approximations are required to reproduce the OD04 scaling.
\item The effect of a optically-thin corrected line-force can be encapsulated in a OTC parameter, which we call \sigsatz{} (\autoref{eq:sat_param}).
Including this does not have much of an effect for O-type stars, since their increased wind density means that there will be less difference in the corrected and uncorrected line-forces.
B-type stars, on the other hand, have their surface mass-flux reduced by approximately 25-30\% when the optically-thin correction is taken into account.
\item The effect of rotation can be similarly represented with a rotation-effect parameter, which we call $\aleph$ (\autoref{eq:rot_param}).
The amount of rotational boosting of the mass-flux is found to depend on both the rotational colatitude and the magnetic obliquity angle.
\item The effects of rotation and the optically-thin correction can not be decoupled, however. We find a different OTC parameter in the case of rotation, \sigsatrot{} (\autoref{eq:sat_rot_param}). Rotation is found to reduce the correction by driving a higher surface mass-flux.
\item The velocity structure within a magnetosphere cannot be described by a global beta-velocity law. 
However, at least for zero rotation, we can well-fit each line with individual beta-velocity laws. 
The best-fit $v_\infty$ and $\beta$ do vary from line to line, however.
With rotation, the beta-velocity law assumption breaks down.
\item The global mass-loss rate for a optically-thin corrected line-force can be accurately estimated by multiplying the optically-thick mass-loss by the OTC parameter, $\Sigma$. We find ``effective'' magnetospheric mass-loss rates, in which the plasma does not fall back to the star, to be approximately 20-65\% of the non-magnetic, non-rotating CAK mass-loss rate.
\end{enumerate}

Overall, we have quantified the effect of a magnetic dipole on a massive star wind with an eye towards better understanding of massive star magnetospheres.
Next steps include adding the finite-disk correction parameter and quantifying its effect on the magnetospheric mass-loss and velocity (Paper II).
Paper III will add colliding wind shocks and the subsequent ``cooling'' region to each line in order to better quantify the level of X-ray emission coming from each line.
This will provide accurate initial conditions for hydrodynamical simulations of centrifugal magnetospheres.

\begin{table*}
\centering
\caption{\label{tab:O_massloss}Same as \autoref{tab:B_massloss}, except for an O-type star with $\eta_* = 100$. All mass-loss rates are given in $10^{-6}~M_\odot$/yr. Numbers in parentheses next to a mass-loss rate represent the ratio of that particular rate to the ``General'' mass-loss with the same rotation.}
\begin{tabular}{lccccccc}
\hline\hline
 &No $B$ & $\omega = 0.0$ & 0.2 & 0.35 & 0.5 & 0.65 & 0.8 \\
\hline
Optically-Thick & 6.61 & 3.77 & 3.79 & 3.83& 3.91 & 4.01 & 4.11\\
General & 6.44 & 3.67 & 3.70  & 3.74& 3.81 & 3.91 & 4.03\\
True & ... & 1.33(0.36) & 1.35(0.36) & 1.35(0.37) & 1.37(0.36) & 1.41(0.36)& 1.44(0.36) \\
Disk & ... & ... & ... & 0.52 (0.14) & 1.12 (0.29) & 1.78(0.46) & 2.56(0.64)\\
Effective & ... & 1.33 (0.36) & 1.35 (0.36) & 1.87(0.5) & 2.49(0.65) & 3.19(0.82) & 3.99 (0.99)\\
\hline
\end{tabular}
\end{table*}

\section*{Acknowledgments}
CB acknowledges support from the NASA GSRP Fellowship (NASA Grant NNX11AK70H). CB and RHDT acknowledge support from NASA ATP Grant NNX12AC72G. The authors thank S. Owocki and J. Cassinelli for helpful discussions and comments.

\appendix

\section{Equation of motion in general case}
\label{app:eom_sat}
Starting with the general \grad{} (\autoref{eq:grad_sat}), we eliminate the density with the continuity equation (\autoref{eq:simp1}):
\begin{align}
\label{eq:g_rad_sat}\grad{}&= \frac{\bar Q\gammael{}GM_*}{1-\alpha}\frac{\psi}{r^2}\frac{|vv'|}{\xi}\left[\left(1+\xi/|vv'|\right)^{1-\alpha} - 1\right],
\end{align}
where we've defined the eigenvalue $\xi \equiv c\kappa_e\bar Q\dot m_* \frac{A_*}{A}$. 
We then get our equation of motion:
\begin{align}
\label{eq:eom_sat}F_\mathrm{gen}&\equiv vv'\left(1 - c_\mathrm{s}^2/v^2\right) - \geff{} \nonumber\\
&~~~~- \frac{\bar Q\gammael{}GM_*}{1-\alpha}\frac{\psi}{r^2}\frac{|vv'|}{\xi}\left[\left(1+\xi/|vv'|\right)^{1-\alpha} - 1\right] = 0,
\end{align}
where, as in the optically-thick case, we ignore the Parker term $c_\mathrm{s}^2\lambda$.

\section{Critical values for optically-thick wind}
\label{app:crit_unsat}
From the equation of motion (\autoref{eq:eom}), the CAK singularity condition (\autoref{eq:gen_sing}), and the CAK regularity condition (\autoref{eq:gen_reg}), we now solve for the critical values ($\Delta_c, u_c$, and $y_c$) as a function of critical point location, \scrit{}. 
For mathematical simplicity, we will define $\Gamma\equiv (A/A_*)^\alpha~\psi/r^2$ in the derivation, such that the starting equation of motion is
\begin{align}
F \equiv y(1-1/u^2) - \geff{} - \Delta\Gamma|y|^\alpha.
\end{align}

We continue with evaluating both CAK critical conditions (\autoref{eq:sing}, \autoref{eq:reg}), remembering that $\geff{}$ and $\Gamma$ are wholly functions of $s$:
\begin{align}
\label{eq:gen_sing}0=\frac{\partial F}{\partial y} = (1-1/u^2) - \alpha\Delta\Gamma|y|^\alpha/y
\end{align}
and
\begin{align}
\label{eq:gen_reg}0=\frac{\partial F}{\partial s} + \frac{y}{c_\mathrm{s}^2u}\frac{\partial F}{\partial u} = -\frac{\partial \geff{}}{\partial s} - \Delta|y|^\alpha\frac{\partial\Gamma}{\partial s} + \frac{2y^2}{c_\mathrm{s}^2u^4}.
\end{align}

Starting from the singularity condition, we get
\begin{align}
\partial F/\partial y = (1-1/u_c^2) - \alpha\Delta_c\Gamma\frac{|y_c|^\alpha}{y_c} = 0,\\
\label{eq:gen_step2}(1-1/u_c^2)y_c = \alpha\Delta_c\Gamma|y_c|^\alpha.
\end{align}
Substituting this into \autoref{eq:mot_unsat_simp} yields
\begin{align}
\label{eq:gen_step3}(1-\alpha)\Delta_c\Gamma|y_c|^\alpha = -\geff{}.
\end{align}
Combining this and \autoref{eq:gen_step2} gives us
\begin{align}
\label{eq:gen_step4}y_c(1-1/u_c^2) = -\frac{\alpha}{1-\alpha}~\geff{}
\end{align}
From the regularity condition, \autoref{eq:gen_reg}, we obtain
\begin{align}
\left(\frac{y_c}{u_c^2}\right)^2 = \frac{c_\mathrm{s}^2}{2}\left[\frac{\partial \geff{}}{\partial s} + \Delta_c|y_c|^\alpha\frac{\partial\Gamma}{\partial s}\right].
\end{align}
Substituting \autoref{eq:gen_step3} into the above yields
\begin{align}
\label{eq:gen_step5}\left(\frac{y_c}{u_c^2}\right)^2 &=\frac{c_\mathrm{s}^2}{2}\Phi^2,
\end{align}
where we define
\begin{align}
\Phi^2 \equiv \frac{\partial \geff{}}{\partial s} - \frac{\geff{}}{1-\alpha}\left(\alpha\lambda+\frac{1}{\psi}\frac{\partial\psi}{\partial s}-\frac{2\psi}{r}\right).
\end{align}

\autoref{eq:gen_step5} has two possible outcomes: $y_c/u_c^2 = \pm c_\mathrm{s}\Phi/\sqrt{2}$.
Since $u_c^2 > 0$, the sign of $y_c$ determines which solution to choose.
For the case of a radiation-driven outflow, the magnetospheric plasma accelerates as it flows away from the stellar surface.
However, due to our sign convention (Section \ref{sec:grid}), outflowing plasma can have either a positive or negative velocity.
For an increasing arc length away from the stellar surface ($\psi > 0$), an accelerating outflow has $v>0$, $dv > 0$, and $ds > 0$, resulting in $y_c > 0$.
At the opposite line footprint ($\psi < 0$), if there is one, an accelerating outflow requires $v<0$, $dv <0$, and $ds<0$, resulting in $y_c < 0$.
Thus, we take the positive (negative) root of \mbox{\autoref{eq:gen_step5}} for positive (negative) $\psi$.

Next, we solve for the critical velocity using \autoref{eq:gen_step4} and \autoref{eq:gen_step5}:
\begin{align}
u_c^2 &= 1\mp\frac{\sqrt{2}\alpha}{(1-\alpha)c_\mathrm{s}\Phi}~\geff{},
\end{align}
which is \autoref{eq:gen_unsat_u}. Substituting \autoref{eq:gen_step4} for $u_c^2$ instead allows us to solve for $y_c$:
\begin{align}
\label{eq:app_yc_unsat}y_c &= \pm\frac{c_\mathrm{s}\Phi}{\sqrt{2}} - \frac{\alpha}{1-\alpha}~\geff{}.
\end{align}
Finally, we solve for $\Delta_c$ using \autoref{eq:gen_step3}:
\begin{align}
\Delta_c &= -\frac{\geff{}}{(1-\alpha)\Gamma|\pm\frac{c_\mathrm{s}\Phi}{\sqrt{2}} - \frac{\alpha}{1-\alpha}~\geff{}|^\alpha}.\nonumber\\
\end{align}

\section{Critical values for general case}
\label{app:crit_sat}
As in Appendix \ref{app:crit_unsat}, we will solve for the critical velocity, velocity derivative, and surface mass-flux as a function of the critical point location. 
First, from the general equation of motion (\autoref{eq:eom_sat}), we make the substitutions $y=vv'$ and $u = v/c_\mathrm{s}$.
Next, for mathematical simplicity, we define $\bar\Gamma = \psi/r^2$:
\begin{align}
\label{eq:eom_sat2}F_\mathrm{gen} &= y(1-1/u^2) - \geff{} \nonumber\\
&~~~~- \frac{\bar Q\gammael{}GM_*}{1-\alpha}\bar\Gamma\frac{|y|}{\xi}\left[\left(1+\xi/|y|\right)^{1-\alpha} - 1\right] = 0.
\end{align}
Next, we evaluate the CAK singularity condition:
\begin{align}
\label{eq:gen_sing_sat}(1-1/u_c^2)y_c &= \frac{\bar Q\gammael{}GM_*}{1-\alpha}\frac{\bar\Gamma}{\xi}|y_c|\left[\left(1+\frac{\xi}{|y_c|}\right)^{1-\alpha} - 1\right]\nonumber\\
&~~~-\bar Q\gammael{}GM_*\bar\Gamma\left(1+\frac{\xi}{|y_c|}\right)^{-\alpha}.
\end{align}
Substituting \autoref{eq:gen_sing_sat} into \autoref{eq:eom_sat2} yields:
\begin{align}
\label{eq:step_sat3}\left(1+\frac{\xi}{|y_c|}\right)^{-\alpha} &= -\frac{\geff{}}{\bar Q\gammael{}GM_*\bar\Gamma}~.
\end{align}
Since the right-hand side of \autoref{eq:step_sat3} is wholly dependent on \scrit{}, we define 
\begin{align}
\chi \equiv \left(1+\xi/|y_c|\right)^{-\alpha} = \left(1+\tau_\mathrm{sob}\right)^{-\alpha},
\end{align} 
$\tau_\mathrm{sob}$ being the Sobolev optical depth.
Additionally, we will define the critical value of $\chi$ as 
\begin{align}
\label{eq:chi_c}\chi_c &= \chi_c(\scrit{})\equiv -\geff{}r^2/[\bar Q\gammael{}GM_*\psi].
\end{align}
We can further simplify this to $\chi_c =\chi_0\aleph_c$, where
\begin{align}
\chi_0 = \frac{1-\gammael{}}{\gammael{}\bar Q}
\end{align}
is the critical $\chi$ value for zero rotation and $\aleph_c$ is the rotation effect parameter (\autoref{eq:rot_param}) evaluated at the critical point.

We discuss the physical meaning of $\chi$ as a ``correction level parameter'' in Section \ref{sec:mflux_sat_zero}, but we note that $\chi_c$ is set by the ratio of the non-radiative external forces to the optically thin ($\tau_\mathrm{sob}\ll 1$) radiative force.
Since $\chi$ can only be between zero and one (since $\tau_\mathrm{sob}>0$), this implies both that, at the critical point, gravity must be stronger than the centrifugal force and the optically thin radiative force must be stronger than the other combined external forces.

These definitions make possible several substitutions:
\begin{subequations}
\begin{align}
\label{eq:gen_step4a}\left(1+\frac{\xi_c}{|y_c|}\right)^{-\alpha} &= \chi_c\\
\left(1+\frac{\xi_c}{|y_c|}\right)^{1-\alpha} &= \chi_c^{\frac{\alpha-1}{\alpha}}\\
\label{eq:gen_step4c}|y_c| = \frac{\xi_c}{\chi_c^{-1/\alpha} - 1}.
\end{align}
\end{subequations}
Now, we evaluate the CAK regularity condition and substitute \autoref{eq:gen_step4a} - \autoref{eq:gen_step4c}:
\begin{align}
\frac{2y_c^2}{c_\mathrm{s}^2u_c^4} &=\frac{\partial \geff{}}{\partial s} -\frac{\geff{}}{1-\alpha}\left[\frac{1-\chi_c^{1/\alpha-1}}{1-\chi_c^{1/\alpha}}\right]\left(\frac{1}{\bar\Gamma}\frac{\partial\bar\Gamma}{\partial s}+ \lambda\right)+\geff{}\lambda,
\end{align}
where we remember $\lambda = \partial A/\partial s / A = -\partial\xi/\partial s/\xi$ (Section \ref{sec:topo}). 
We can then define $\bar\Phi$ in a parallel manner to the optically-thick case ($\Phi$; \autoref{eq:gen_phi_unsat}):
\begin{align}
\label{eq:gen_phi_sat}\bar\Phi^2 &\equiv\frac{\partial \geff{}}{\partial s} -  \frac{\geff{}}{1-\alpha}\left[\frac{1-\chi_c^{1/\alpha-1}}{1-\chi_c^{1/\alpha}}\right]\left(\frac{1}{\psi}\frac{\partial\psi}{\partial s} - \frac{2\psi}{r}\right) \nonumber\\
&~~~~+ \frac{1-\alpha-\left[\frac{1-\chi_c^{1/\alpha-1}}{1-\chi_c^{1/\alpha}}\right]}{1-\alpha}\geff{}\lambda
\end{align}
As a check, we note that the general case reduces to the optically-thick case for $\xi/|y_c| \gg 1$ (i.e. $\tau_\mathrm{sob} \gg 1$). 
This leads to $\chi_c \ll 1$; the terms in the brackets above reduce to 1.
From this, it is easily seen that $\bar\Phi^2\to\Phi^2$.

We can use \autoref{eq:eom_sat2} and $\bar\Phi^2 = 2y_c^2/(c_s^2u_c^4)$ to solve for $y_c$, remembering that both roots of \autoref{eq:gen_phi_sat} are valid solutions for an accelerating outflow from the stellar surface (c.f. Section \ref{sec:topo}):
\begin{align}
\label{eq:gen_y_c_sat}y_c &= \frac{1-\alpha-\left[\frac{1-\chi_c^{1/\alpha-1}}{1-\chi_c^{1/\alpha}}\right]}{1-\alpha}~\geff{}\pm\frac{c_\mathrm{s}\bar\Phi}{\sqrt{2}}
\end{align}
where the top (bottom) term applies for positive (negative) $\psi$. 
In the optically-thick limit, the term in brackets is $\approx 1$ and the overall equation reduces to the optically-thick $y_c$ (\autoref{eq:gen_unsat_y}).
Similarly, we can obtain the critical velocity:
\begin{align}
\label{eq:gen_u_c_sat}u_c^2 &= 1\pm\frac{\sqrt{2}}{c_\mathrm{s}\bar\Phi}\left(\frac{1-\alpha-\left[\frac{1-\chi_c^{1/\alpha-1}}{1-\chi_c^{1/\alpha}}\right]}{1-\alpha}\right)~\geff{}
\end{align}
which reduces to the optically-thick value \autoref{eq:gen_unsat_u} in the proper limit. 
Finally, we solve for the critical mass-flux using \autoref{eq:gen_step4c}:
\begin{align}
\label{eq:gen_sat_mflux}\dot m_* &= \frac{(\chi_c^{-1/\alpha} - 1)|y_c|}{c\kappa_e\bar Q\frac{A_*}{A}}
\end{align}

As in the optically-thick case, these critical values imply a range to the allowable critical point location, \scrit{}, by requiring that $\bar\Phi^2 > 0$ and $\chi_c > 0$.
With a boundary condition (e.g. \autoref{eq:boundary_v}), we can solve for \scrit{} iteratively using the procedure described in Section \ref{sec:crit}.

\section{Dipole parameterizations}
\label{app:dip}
There are several spatial variables that we need to translate from $r$ or $s$ to the magnetic colatitude $\tilde\theta$, which we will do in the following paragraphs.
For an aligned dipole, the magnetic axis coincides with the rotational axis, so we take $\tilde\theta = \theta$.
Since the plasma flows along the field line but most of our external forces (Section \ref{sec:Euler}) depend on the radial distance, it is convenient to simplify the derivative along the field line as $\partial/\partial s = \partial r/\partial s~\partial/\partial r$.
Towards this end, we calculate $\partial r/\partial s$ from \autoref{eq:r_theta} and \autoref{eq:ds_theta}:
\begin{subequations}
\begin{align}
\frac{\partial r}{\partial\theta} &= 2r_m\sin\theta\cos\theta\\
\frac{\partial s}{\partial\theta} &= r_m\sin\theta\sqrt{1+3\cos^2\theta}\\
\frac{\partial r}{\partial s} &= \frac{2\cos\theta}{\sqrt{1+3\cos^2\theta}}
\end{align}
\end{subequations}
We also have $\psi = \hat r\cdot\hat s = \partial r/\partial s$. Next, from the the conservation of magnetic flux ($BA~=$ const) and our magnetic field definition \autoref{eq:dipole_B}:
\begin{align}
\label{eq:dipole_A_star_A}\frac{A_*}{A} &= \left(\frac{R_*}{r}\right)^3\frac{\sqrt{1+3\cos^2\theta}}{\sqrt{1+3\cos^2\theta_m}},
\end{align}
where the surface radius $R_*$ is defined in \autoref{eq:R_surf}. From this, we simplify $\lambda = \partial A/\partial s/A$:
\begin{align}
\label{eq:dipole_lambda}\lambda &= \psi\left[\frac{3}{r}+\frac{3}{2r_m(1+3\cos^2\theta)}\right].
\end{align}

Now, we move on to parameterizing the external forces on the plasma in the magnetosphere. From Section \ref{sec:geff}, we get:
\begin{align}
\label{eq:dipole_C}\geff{} + c_\mathrm{s}^2 \lambda&=-(1-\gammael{})\frac{GM_*\psi}{r^2}\left(1 - \frac{8\omega^2}{27R_p^3}\frac{\bar r r^2\bar\psi}{\psi}\right) + c_\mathrm{s}^2\lambda,
\end{align}
where $\bar r = r \sin\theta$ is the distance from the rotational axis and $\bar \psi = \hat{\bar r} \cdot \hat s$. Typically, the Parker term $c_\mathrm{s}^2\lambda$ is neglected.
Parameterizing $\bar r$ and $\bar\psi$ yields
\begin{align}
\bar r &= r\sin\theta = r_m \sin^3{\theta}~,\\
\frac{\partial\bar r}{\partial\theta} &= 3r_m\sin^2\theta\cos\theta~,\\
\bar\psi &= \frac{\partial\bar r}{\partial s} = \frac{3\sin\theta\cos\theta}{\sqrt{1+3\cos^2\theta}}~.
\end{align}
Since there are several spatial derivatives in the critical point calculations, we calculate $\partial/\partial s$ of several variables:
\begin{align}
\frac{\partial\psi}{\partial s} &= \frac{-2}{r_m(1+3\cos^2\theta)^2},\\
\frac{\partial\bar\psi}{\partial s} &=\frac{3(\cos^2\theta-\sin^2\theta) + \bar\psi^2}{r_m\sin\theta(1+3\cos^2\theta)},\\
\frac{\partial\lambda}{\partial s} &= \frac{\partial\psi}{\partial s}\left[\frac{3}{r}+\frac{3}{2r_m(1+3\cos^2\theta)}\right] \nonumber\\
&~~~~~- \frac{3\psi^2}{r^2} + \frac{9\psi\cos\theta}{r_m^2(1+3\cos^2\theta)^{5/2}}.
\end{align}
Now, we present $\partial \geff{}/\partial s$ as required for $\Phi$ in (\ref{eq:gen_phi_unsat}):
\begin{align}
\frac{\partial \geff{}}{\partial s} &= (1-\gammael{})GM_*\left[\frac{2\psi^2}{r^3} - \frac{\partial\psi/\partial s}{r^2} + \frac{8\omega^2}{27R_p^3}\left(\bar\psi^2 + \bar r\frac{\partial\bar\psi}{\partial s}\right)\right].
\end{align}

With all our definitions above, we can easily solve for the critical point values (\autoref{eq:gen_unsat_u} - \autoref{eq:gen_unsat_Delta}), and the critical surface mass-flux (\autoref{eq:gen_unsat_mflux}) for a magnetic dipole.
In theory, one could write the full critical value equations wholly in terms of $\theta$, but such a representation would be too muddled and provide no benefit.
For practical purposes, it is much easier to calculate values along each input magnetic field line and derive the spatial derivatives numerically.

The procedure for calculating the dipole critical values for a general line acceleration is identical to above, since the spatial variables are the same.

\section{Mass-flux scaling for general case}
\label{app:mflux_scaling_sat}
We reproduce the scaling of Section \ref{sec:mflux_zero} for a general line acceleration.
As in the optically-thick case, we ignore both the Parker term $c_\mathrm{s}^2\lambda$ and $\Phi$:
\begin{align}
y_c&\approx \frac{1-\alpha-\left[\frac{1-\chi_c^{1/\alpha-1}}{1-\chi_c^{1/\alpha}}\right]}{1-\alpha}~\geff{} = -\frac{\alpha}{1-\alpha}\sigsat{}\geff{},
\end{align}
with $\chi_c$ defined in \autoref{eq:chi_c} and we define a optically-thin correction parameter
\begin{align}
\sigsat{} \equiv \frac{\left|1-\alpha-\left[\frac{1-\chi_c^{1/\alpha-1}}{1-\chi_c^{1/\alpha}}\right]\right|}{\alpha},
\end{align}
whose utility will become evident later on.

Next, we simplify \autoref{eq:gen_sat_mflux} for an aligned dipole, noting that usually $\chi_c^{-1/\alpha}\gg 1$:
\begin{align}
\label{eq:mflux_approx_sat}\dot m_* &\approx \frac{\alpha}{1-\alpha}\sigsat{}\frac{L_*}{4\pi c^2} |\geff{}|\left[\frac{(\bar Q\gammael{}GM_*)^{1-\alpha}}{-\geff{}}\right]^{1/\alpha}\nonumber\\
&~~~~\times \frac{\psi_c^{1/\alpha}r^{3-2/\alpha}_c}{R_*^3}\sqrt{\frac{1+3\cos^2\theta_m}{1+3\cos^2\theta_c}}
\end{align}
which is identical to \autoref{eq:mflux_approx_unsat} except for the OTC parameter.
For zero rotation, $\geff{} = -(1-\Gamma_\mathrm{el})GM_*\psi/r^2$ and $R_* = R_p$:
\begin{align}
\label{eq:mdip_sat_app}\dot m_* \approx\mdip{}\sigsatz{}\psi_c\sqrt{\frac{1+3\cos^2\theta_m}{1+3\cos^2\theta_c}},
\end{align}
with \mdip{}, the optically-thick surface mass-flux, defined in \autoref{eq:mdip}.
Here, \sigsatz{} is the OTC parameter for zero rotation (as discussed in Section \ref{sec:mflux_sat_zero}).

Taking the OD04-type simplifications $\rcritz{} - R_* \ll R_*$, $\theta_c\approx\theta_m$ and $\psi_c\approx\psi_m = \mu_B$ (for zero rotation), we get
\begin{align}
\label{eq:flux_scaling_sat_app}\dot m_* \approx \mu_B\sigsatz{}\mdip{}(\rcritz{}),
\end{align}
which is nearly identical to \autoref{eq:flux_scaling_unsat} with the addition of the OTC parameter.
$\rcritz{}$ is different between the general and optically-thick cases (Section \ref{sec:rcrit}), however, so that should be taken into account.

\bibliography{AMCWS_bib}

\begin{thebibliography}{}
\makeatletter
\relax
\def\mn@urlcharsother{\let\do\@makeother \do\$\do\&\do\#\do\^\do\_\do\%\do\~}
\def\mn@doi{\begingroup\mn@urlcharsother \@ifnextchar [ {\mn@doi@}
  {\mn@doi@[]}}
\def\mn@doi@[#1]#2{\def\@tempa{#1}\ifx\@tempa\@empty \href
  {http://dx.doi.org/#2} {doi:#2}\else \href {http://dx.doi.org/#2} {#1}\fi
  \endgroup}
\def\mn@eprint#1#2{\mn@eprint@#1:#2::\@nil}
\def\mn@eprint@arXiv#1{\href {http://arxiv.org/abs/#1} {{\tt arXiv:#1}}}
\def\mn@eprint@dblp#1{\href {http://dblp.uni-trier.de/rec/bibtex/#1.xml}
  {dblp:#1}}
\def\mn@eprint@#1:#2:#3:#4\@nil{\def\@tempa {#1}\def\@tempb {#2}\def\@tempc
  {#3}\ifx \@tempc \@empty \let \@tempc \@tempb \let \@tempb \@tempa \fi \ifx
  \@tempb \@empty \def\@tempb {arXiv}\fi \@ifundefined
  {mn@eprint@\@tempb}{\@tempb:\@tempc}{\expandafter \expandafter \csname
  mn@eprint@\@tempb\endcsname \expandafter{\@tempc}}}

\bibitem[\protect\citeauthoryear{{Abbott}}{{Abbott}}{1980}]{abbott80}
{Abbott} D.~C.,  1980, \mn@doi [\apj] {10.1086/158550}, \href
  {http://adsabs.harvard.edu/abs/1980ApJ...242.1183A} {242, 1183}

\bibitem[\protect\citeauthoryear{{Abbott}}{{Abbott}}{1982}]{abbott82}
{Abbott} D.~C.,  1982, \mn@doi [\apj] {10.1086/160166}, \href
  {http://adsabs.harvard.edu/abs/1982ApJ...259..282A} {259, 282}

\bibitem[\protect\citeauthoryear{{Babel} \& {Montmerle}}{{Babel} \&
  {Montmerle}}{1997}]{babel97}
{Babel} J.,  {Montmerle} T.,  1997, \aap, \href
  {http://adsabs.harvard.edu/abs/1997A\%26A...323..121B} {323, 121}

\bibitem[\protect\citeauthoryear{{Bjorkman}}{{Bjorkman}}{1995}]{bjorkman95}
{Bjorkman} J.~E.,  1995, \mn@doi [\apj] {10.1086/176396}, \href
  {http://adsabs.harvard.edu/abs/1995ApJ...453..369B} {453, 369}

\bibitem[\protect\citeauthoryear{{Bohlender} \& {Monin}}{{Bohlender} \&
  {Monin}}{2011}]{bohlender11}
{Bohlender} D.~A.,  {Monin} D.,  2011, \mn@doi [\aj]
  {10.1088/0004-6256/141/5/169}, \href
  {http://adsabs.harvard.edu/abs/2011AJ....141..169B} {141, 169}

\bibitem[\protect\citeauthoryear{{Cassinelli}}{{Cassinelli}}{1979}]{cassinelli79}
{Cassinelli} J.~P.,  1979, \mn@doi [\araa]
  {10.1146/annurev.aa.17.090179.001423}, \href
  {http://adsabs.harvard.edu/abs/1979ARA\%26A..17..275C} {17, 275}

\bibitem[\protect\citeauthoryear{{Castor}, {Abbott}  \& {Klein}}{{Castor}
  et~al.}{1975}]{cak75}
{Castor} J.~I.,  {Abbott} D.~C.,   {Klein} R.~I.,  1975, \mn@doi [\apj]
  {10.1086/153315}, \href {http://adsabs.harvard.edu/abs/1975ApJ...195..157C}
  {195, 157}

\bibitem[\protect\citeauthoryear{{Chandra} et~al.,}{{Chandra}
  et~al.}{2015}]{chandra15}
{Chandra} P.,  et~al., 2015, \mn@doi [\mnras] {10.1093/mnras/stv1378}, \href
  {http://adsabs.harvard.edu/abs/2015MNRAS.452.1245C} {452, 1245}

\bibitem[\protect\citeauthoryear{{Cranmer} \& {Owocki}}{{Cranmer} \&
  {Owocki}}{1996}]{cranmer96a}
{Cranmer} S.~R.,  {Owocki} S.~P.,  1996, \mn@doi [\apj] {10.1086/177166}, \href
  {http://adsabs.harvard.edu/abs/1996ApJ...462..469C} {462, 469}

\bibitem[\protect\citeauthoryear{{Cur{\'e}} \& {Rial}}{{Cur{\'e}} \&
  {Rial}}{2004}]{cure04}
{Cur{\'e}} M.,  {Rial} D.~F.,  2004, \mn@doi [\aap]
  {10.1051/0004-6361:20040325}, \href
  {http://adsabs.harvard.edu/abs/2004A\%26A...428..545C} {428, 545}

\bibitem[\protect\citeauthoryear{{Donati} et~al.,}{{Donati}
  et~al.}{2006}]{donati06}
{Donati} J.-F.,  et~al., 2006, \mn@doi [\mnras]
  {10.1111/j.1365-2966.2006.10558.x}, \href
  {http://adsabs.harvard.edu/abs/2006MNRAS.370..629D} {370, 629}

\bibitem[\protect\citeauthoryear{{Drew}}{{Drew}}{1989}]{drew89}
{Drew} J.~E.,  1989, \mn@doi [\apjs] {10.1086/191374}, \href
  {http://adsabs.harvard.edu/abs/1989ApJS...71..267D} {71, 267}

\bibitem[\protect\citeauthoryear{{Eikenberry} et~al.,}{{Eikenberry}
  et~al.}{2014}]{eikenberry14}
{Eikenberry} S.~S.,  et~al., 2014, \mn@doi [\apjl]
  {10.1088/2041-8205/784/2/L30}, \href
  {http://adsabs.harvard.edu/abs/2014ApJ...784L..30E} {784, L30}

\bibitem[\protect\citeauthoryear{{Friend} \& {Abbott}}{{Friend} \&
  {Abbott}}{1986}]{friend86}
{Friend} D.~B.,  {Abbott} D.~C.,  1986, \mn@doi [\apj] {10.1086/164809}, \href
  {http://adsabs.harvard.edu/abs/1986ApJ...311..701F} {311, 701}

\bibitem[\protect\citeauthoryear{{Gayley}}{{Gayley}}{1995}]{gayley95}
{Gayley} K.~G.,  1995, \mn@doi [\apj] {10.1086/176492}, \href
  {http://adsabs.harvard.edu/abs/1995ApJ...454..410G} {454, 410}

\bibitem[\protect\citeauthoryear{{Gayley} \& {Owocki}}{{Gayley} \&
  {Owocki}}{2000}]{gayley00}
{Gayley} K.~G.,  {Owocki} S.~P.,  2000, \mn@doi [\apj] {10.1086/309002}, \href
  {http://adsabs.harvard.edu/abs/2000ApJ...537..461G} {537, 461}

\bibitem[\protect\citeauthoryear{{Grunhut} et~al.,}{{Grunhut}
  et~al.}{2012}]{grunhut12}
{Grunhut} J.~H.,  et~al., 2012, \mn@doi [\mnras]
  {10.1111/j.1365-2966.2011.19824.x}, \href
  {http://adsabs.harvard.edu/abs/2012MNRAS.419.1610G} {419, 1610}

\bibitem[\protect\citeauthoryear{{Howarth} et~al.,}{{Howarth}
  et~al.}{2007}]{howarth07}
{Howarth} I.~D.,  et~al., 2007, \mn@doi [\mnras]
  {10.1111/j.1365-2966.2007.12178.x}, \href
  {http://adsabs.harvard.edu/abs/2007MNRAS.381..433H} {381, 433}

\bibitem[\protect\citeauthoryear{{Kochukhov}, {Lundin}, {Romanyuk}  \&
  {Kudryavtsev}}{{Kochukhov} et~al.}{2011}]{kochukhov11}
{Kochukhov} O.,  {Lundin} A.,  {Romanyuk} I.,   {Kudryavtsev} D.,  2011,
  \mn@doi [\apj] {10.1088/0004-637X/726/1/24}, \href
  {http://adsabs.harvard.edu/abs/2011ApJ...726...24K} {726, 24}

\bibitem[\protect\citeauthoryear{{Kudritzki}}{{Kudritzki}}{2002}]{kudritzki02}
{Kudritzki} R.~P.,  2002, \mn@doi [\apj] {10.1086/342178}, \href
  {http://adsabs.harvard.edu/abs/2002ApJ...577..389K} {577, 389}

\bibitem[\protect\citeauthoryear{{Kudritzki} \& {Puls}}{{Kudritzki} \&
  {Puls}}{2000}]{kudritzki00}
{Kudritzki} R.-P.,  {Puls} J.,  2000, \mn@doi [\araa]
  {10.1146/annurev.astro.38.1.613}, \href
  {http://adsabs.harvard.edu/abs/2000ARA\%26A..38..613K} {38, 613}

\bibitem[\protect\citeauthoryear{{Kudritzki}, {Pauldrach}, {Puls}  \&
  {Abbott}}{{Kudritzki} et~al.}{1989}]{kudritzki89}
{Kudritzki} R.~P.,  {Pauldrach} A.,  {Puls} J.,   {Abbott} D.~C.,  1989, \aap,
  \href {http://adsabs.harvard.edu/abs/1989A\%26A...219..205K} {219, 205}

\bibitem[\protect\citeauthoryear{{Lamers} \& {Cassinelli}}{{Lamers} \&
  {Cassinelli}}{1999}]{lamers99}
{Lamers} H.~J.~G.~L.~M.,  {Cassinelli} J.~P.,  1999, {Introduction to Stellar
  Winds}

\bibitem[\protect\citeauthoryear{{Linsky}, {Drake}  \& {Bastian}}{{Linsky}
  et~al.}{1992}]{linsky92}
{Linsky} J.~L.,  {Drake} S.~A.,   {Bastian} T.~S.,  1992, \mn@doi [\apj]
  {10.1086/171509}, \href {http://adsabs.harvard.edu/abs/1992ApJ...393..341L}
  {393, 341}

\bibitem[\protect\citeauthoryear{{Lucy}}{{Lucy}}{2007}]{lucy07}
{Lucy} L.~B.,  2007, \mn@doi [\aap] {10.1051/0004-6361:20078236}, \href
  {http://adsabs.harvard.edu/abs/2007A\%26A...474..701L} {474, 701}

\bibitem[\protect\citeauthoryear{{Madura}, {Owocki}  \& {Feldmeier}}{{Madura}
  et~al.}{2007}]{madura07}
{Madura} T.~I.,  {Owocki} S.~P.,   {Feldmeier} A.,  2007, \mn@doi [\apj]
  {10.1086/512602}, \href {http://adsabs.harvard.edu/abs/2007ApJ...660..687M}
  {660, 687}

\bibitem[\protect\citeauthoryear{{Marlborough} \& {Zamir}}{{Marlborough} \&
  {Zamir}}{1984}]{marlborough84}
{Marlborough} J.~M.,  {Zamir} M.,  1984, \mn@doi [\apj] {10.1086/161657}, \href
  {http://adsabs.harvard.edu/abs/1984ApJ...276..706M} {276, 706}

\bibitem[\protect\citeauthoryear{{Morel} et~al.,}{{Morel}
  et~al.}{2015}]{morel15}
{Morel} T.,  et~al., 2015, in {Meynet} G.,  {Georgy} C.,  {Groh} J.,   {Stee}
  P.,  eds,  IAU Symposium Vol. 307, IAU Symposium. pp 342--347 (\mn@eprint
  {arXiv} {1408.2100}), \mn@doi{10.1017/S1743921314007054}

\bibitem[\protect\citeauthoryear{{M{\"u}ller} \& {Vink}}{{M{\"u}ller} \&
  {Vink}}{2008}]{mueller08}
{M{\"u}ller} P.~E.,  {Vink} J.~S.,  2008, \mn@doi [\aap]
  {10.1051/0004-6361:20078798}, \href
  {http://adsabs.harvard.edu/abs/2008A\%26A...492..493M} {492, 493}

\bibitem[\protect\citeauthoryear{{Naz{\'e}}, {Petit}, {Rinbrand}, {Cohen},
  {Owocki}, {ud-Doula}  \& {Wade}}{{Naz{\'e}} et~al.}{2014}]{naze14}
{Naz{\'e}} Y.,  {Petit} V.,  {Rinbrand} M.,  {Cohen} D.,  {Owocki} S.,
  {ud-Doula} A.,   {Wade} G.~A.,  2014, \mn@doi [\apjs]
  {10.1088/0067-0049/215/1/10}, \href
  {http://adsabs.harvard.edu/abs/2014ApJS..215...10N} {215, 10}

\bibitem[\protect\citeauthoryear{{Naz{\'e}}, {Sundqvist}, {Fullerton},
  {ud-Doula}, {Wade}, {Rauw}  \& {Walborn}}{{Naz{\'e}} et~al.}{2015}]{naze15}
{Naz{\'e}} Y.,  {Sundqvist} J.~O.,  {Fullerton} A.~W.,  {ud-Doula} A.,  {Wade}
  G.~A.,  {Rauw} G.,   {Walborn} N.~R.,  2015, \mn@doi [\mnras]
  {10.1093/mnras/stv1445}, \href
  {http://adsabs.harvard.edu/abs/2015MNRAS.452.2641N} {452, 2641}

\bibitem[\protect\citeauthoryear{{Noebauer} \& {Sim}}{{Noebauer} \&
  {Sim}}{2015}]{noebauer15}
{Noebauer} U.~M.,  {Sim} S.~A.,  2015, \mn@doi [\mnras]
  {10.1093/mnras/stv1849}, \href
  {http://adsabs.harvard.edu/abs/2015MNRAS.453.3120N} {453, 3120}

\bibitem[\protect\citeauthoryear{{Owocki} \& {ud-Doula}}{{Owocki} \&
  {ud-Doula}}{2004}]{owocki04}
{Owocki} S.~P.,  {ud-Doula} A.,  2004, \mn@doi [\apj] {10.1086/380123}, \href
  {http://adsabs.harvard.edu/abs/2004ApJ...600.1004O} {600, 1004}

\bibitem[\protect\citeauthoryear{{Owocki}, {Castor}  \& {Rybicki}}{{Owocki}
  et~al.}{1988}]{owocki88}
{Owocki} S.~P.,  {Castor} J.~I.,   {Rybicki} G.~B.,  1988, \mn@doi [\apj]
  {10.1086/166977}, \href {http://adsabs.harvard.edu/abs/1988ApJ...335..914O}
  {335, 914}

\bibitem[\protect\citeauthoryear{{Pauldrach}, {Puls}  \&
  {Kudritzki}}{{Pauldrach} et~al.}{1986}]{pauldrach86}
{Pauldrach} A.,  {Puls} J.,   {Kudritzki} R.~P.,  1986, \aap, \href
  {http://adsabs.harvard.edu/abs/1986A\%26A...164...86P} {164, 86}

\bibitem[\protect\citeauthoryear{{Petit} et~al.,}{{Petit}
  et~al.}{2013}]{petit13}
{Petit} V.,  et~al., 2013, \mn@doi [\mnras] {10.1093/mnras/sts344}, \href
  {http://adsabs.harvard.edu/abs/2013MNRAS.429..398P} {429, 398}

\bibitem[\protect\citeauthoryear{{Puls}, {Springmann}  \& {Lennon}}{{Puls}
  et~al.}{2000}]{puls00}
{Puls} J.,  {Springmann} U.,   {Lennon} M.,  2000, \mn@doi [\aaps]
  {10.1051/aas:2000312}, \href
  {http://adsabs.harvard.edu/abs/2000A\%26AS..141...23P} {141, 23}

\bibitem[\protect\citeauthoryear{{Schure}, {Kosenko}, {Kaastra}, {Keppens}  \&
  {Vink}}{{Schure} et~al.}{2009}]{schure09}
{Schure} K.~M.,  {Kosenko} D.,  {Kaastra} J.~S.,  {Keppens} R.,   {Vink} J.,
  2009, \mn@doi [\aap] {10.1051/0004-6361/200912495}, \href
  {http://adsabs.harvard.edu/abs/2009A\%26A...508..751S} {508, 751}

\bibitem[\protect\citeauthoryear{{Sundqvist} \& {Owocki}}{{Sundqvist} \&
  {Owocki}}{2015}]{sundqvist15}
{Sundqvist} J.~O.,  {Owocki} S.~P.,  2015, \mn@doi [\mnras]
  {10.1093/mnras/stv1858}, \href
  {http://adsabs.harvard.edu/abs/2015MNRAS.453.3428S} {453, 3428}

\bibitem[\protect\citeauthoryear{{Townsend} \& {Owocki}}{{Townsend} \&
  {Owocki}}{2005}]{townsend05}
{Townsend} R.~H.~D.,  {Owocki} S.~P.,  2005, \mnras, 357, 251

\bibitem[\protect\citeauthoryear{{Townsend}, {Owocki}  \&
  {Ud-Doula}}{{Townsend} et~al.}{2007}]{townsend07}
{Townsend} R.~H.~D.,  {Owocki} S.~P.,   {Ud-Doula} A.,  2007, \mn@doi [\mnras]
  {10.1111/j.1365-2966.2007.12427.x}, \href
  {http://adsabs.harvard.edu/abs/2007MNRAS.382..139T} {382, 139}

\bibitem[\protect\citeauthoryear{{Townsend} et~al.,}{{Townsend}
  et~al.}{2013}]{townsend13}
{Townsend} R.~H.~D.,  et~al., 2013, \mn@doi [\apj]
  {10.1088/0004-637X/769/1/33}, \href
  {http://adsabs.harvard.edu/abs/2013ApJ...769...33T} {769, 33}

\bibitem[\protect\citeauthoryear{{Vink}, {de Koter}  \& {Lamers}}{{Vink}
  et~al.}{2000}]{vink00}
{Vink} J.~S.,  {de Koter} A.,   {Lamers} H.~J.~G.~L.~M.,  2000, \aap, \href
  {http://adsabs.harvard.edu/abs/2000A\%26A...362..295V} {362, 295}

\bibitem[\protect\citeauthoryear{{Wade} et~al.,}{{Wade} et~al.}{2014}]{wade14}
{Wade} G.~A.,  et~al., 2014, in IAU Symposium. pp 265--269,
  \mn@doi{10.1017/S1743921314002233}

\bibitem[\protect\citeauthoryear{{White} \& {Chen}}{{White} \&
  {Chen}}{1995}]{white95}
{White} R.~L.,  {Chen} W.,  1995, in {van der Hucht} K.~A.,  {Williams} P.~M.,
  eds,  IAU Symposium Vol. 163, Wolf-Rayet Stars: Binaries; Colliding Winds;
  Evolution. p.~438

\bibitem[\protect\citeauthoryear{{ud-Doula} \& {Owocki}}{{ud-Doula} \&
  {Owocki}}{2002}]{uddoula02}
{ud-Doula} A.,  {Owocki} S.~P.,  2002, \mn@doi [\apj] {10.1086/341543}, \href
  {http://adsabs.harvard.edu/abs/2002ApJ...576..413U} {576, 413}

\bibitem[\protect\citeauthoryear{{ud-Doula}, {Owocki}  \&
  {Townsend}}{{ud-Doula} et~al.}{2008}]{uddoula08}
{ud-Doula} A.,  {Owocki} S.~P.,   {Townsend} R.~H.~D.,  2008, \mn@doi [\mnras]
  {10.1111/j.1365-2966.2008.12840.x}, \href
  {http://adsabs.harvard.edu/abs/2008MNRAS.385...97U} {385, 97}

\bibitem[\protect\citeauthoryear{{ud-Doula}, {Owocki}  \&
  {Townsend}}{{ud-Doula} et~al.}{2009}]{uddoula09}
{ud-Doula} A.,  {Owocki} S.~P.,   {Townsend} R.~H.~D.,  2009, \mn@doi [\mnras]
  {10.1111/j.1365-2966.2008.14134.x}, \href
  {http://adsabs.harvard.edu/abs/2009MNRAS.392.1022U} {392, 1022}

\bibitem[\protect\citeauthoryear{{ud-Doula}, {Owocki}, {Townsend}, {Petit}  \&
  {Cohen}}{{ud-Doula} et~al.}{2014}]{uddoula14}
{ud-Doula} A.,  {Owocki} S.,  {Townsend} R.,  {Petit} V.,   {Cohen} D.,  2014,
  \mn@doi [\mnras] {10.1093/mnras/stu769}, \href
  {http://adsabs.harvard.edu/abs/2014MNRAS.441.3600U} {441, 3600}

\makeatother
\end{thebibliography}

\label{lastpage}

\end{document}